# Dynamical phase-field model of cavity electromagnonic systems

Shihao Zhuang,[1a)] Yujie Zhu,[1a)] Changchun Zhong,[2] Liang Jiang,[2] Xufeng Zhang[3,4†], Jia-Mian Hu[1*]

[1]Department of Materials Science and Engineering, University of Wisconsin-Madison, Madison, WI, 53706, USA

[2]Pritzker School of Molecular Engineering, The University of Chicago, Chicago, IL 60637, USA

[3]Department of Electrical and Computer Engineering, Northeastern University, Boston, Massachusetts 02115, USA

[4]Department of Physics, Northeastern University, Boston, MA 02115, USA

## Abstract

Cavity electromagnonic system, which simultaneously consists of cavities for photons, magnons (quanta of spin waves), and acoustic phonons, provides an exciting platform to achieve coherent energy transduction among different physical systems down to single quantum level. Here we report a dynamical phase-field model that allows simulating the coupled dynamics of the electromagnetic waves, magnetization, and strain in 3D multiphase systems. As examples of application, we computationally demonstrate the excitation of hybrid magnon-photon modes (magnon polaritons), Floquet-induced magnonic Aulter-Townes splitting, dynamical energy exchange (Rabi oscillation) and relative phase control (Ramsey interference) between the two magnon polariton modes. The simulation results are consistent with analytical calculations based on Floquet Hamiltonian theory. Simulations are also performed to design a cavity electro-magno-mechanical system that enables the triple phonon-magnon-photon resonance, where the resonant excitation of a chiral, fundamental ($n$=1) transverse acoustic phonon mode by magnon polaritons is demonstrated. With the capability to predict coupling strength, dissipation rates, and temporal evolution of photon/magnon/phonon mode profiles using fundamental materials parameters as the inputs, the present dynamical phase-field model represents a valuable computational tool to guide the fabrication of the cavity electromagnonic system and the design of operating conditions for applications in quantum sensing, transduction, and communication.

E-mails:
[†]xu.zhang@northeastern.edu
[*]jhu238@wisc.edu

[a)]S.Z. and Y.Z. contribute equally to this work.

**Introduction**

One main goal of the cavity electromagnonics is to realize strong and dynamically tunable coupling between magnons (quanta of spin waves) and cavity photons (quanta of confined electromagnetic waves)[1–3], with application potential in quantum storage[4,5], quantum transduction[6,7] and quantum sensing[8]. The strong coupling between the Kittel mode magnon (spatially uniform precession of magnetization) and the cavity photon was theoretically predicted by Soykal and Flatté[9,10] and experimentally observed in hybrid systems that involve yttrium iron garnet (YIG) bulk crystals[11,12], permalloy thin-film stripe[13], YIG film[14,15] mounted on a coplanar microwave resonator, or YIG bulk crystal sphere(s)/slab inside a three-dimensional (3D) microwave cavity[4,16–25]. One key feature of such strong coupling is the mode frequency splitting with an avoided crossing in the frequency spectrum, which indicates the hybridization of magnon and photon into a new quasiparticle called magnon polariton[14,19,26]. In the time-domain, the energy of magnon polaritons is constantly exchanged between the magnon and the photon system with 100% conversion efficiency.

To realize practical quantum operation such as mode swapping and storage[27], it is necessary to dynamically control the exchange process between the two hybrid modes of magnon polaritons upon the completion of transferring a single quantum of excitation[2]. For example, Floquet engineering[28] — which herein refers to the simultaneous application of a periodic driving magnetic field — has been successfully implemented to *in situ* control the transition between the two hybrid modes and even induce further splitting of each mode into two energy levels associated with different Floquet modes[21], analogous to the Autler-Townes splitting in atomic physics.

In addition to the studies on magnon-photon resonance, tripartite coupling among the photons, magnons, and phonons have also been demonstrated experimentally in a cavity electromagnonic system[15,22–25], which can also be called a cavity electro-magno-mechanical system in this case. For example, Zhang *et al* reported a resonant coupling among the Kittel mode magnons, cavity photons, and high-overtone bulk acoustic phonons — all having the same frequency of a few gigahertz (GHz) — in a $Gd_3Ga_5O_{12}$(GGG, substrate)/YIG(film, 200-nm-thick) mounted on a split-ring resonator[15]. In a 0.25-mm-diameter YIG sphere placed in a 3D photon cavity, Zhang *et al* [22] demonstrated a coherent coupling between a GHz magnon polariton (with a frequency $\omega_+$ or $\omega_-$) and a megahertz (MHz) acoustic phonon (frequency: $\omega_p$) by parametrically driving the cavity with a strong microwave signal at a frequency $\omega_d$, with $\omega_d-\omega_-=-\omega_p$ or $\omega_d-\omega_+=\omega_p$.

The main objective of this article is to report a 3D dynamical phase-field model that enables simulating and predicting the coupled dynamics of photons, magnons, and acoustic phonons in a cavity electromagnonic system comprised of a magnon/phonon resonator placed in a bulk 3D photon cavity, which is one of the most used structures in experiments[4,16–25]. In contrast to the fact that Hamiltonian-based theoretical analyses (e.g., [21,29]) need to take the mode coupling strength as the input and are therefore not predictive, the present dynamic phase-field model allows for predicting the spatiotemporal evolution of coupled modes in 3D photon cavity and magnon resonators of arbitrary size and geometry under various operation conditions, using only the fundamental materials parameters as the input. Therefore, it can be used to guide the design of cavity structure and control conditions for realizing desirable quantum operation.

The dynamically evolving physical parameters in a cavity electromagnonic system include the magnetic-field component ($\mathbf{H}^{EM}$) of the electromagnetic (EM) wave in the microwave cavity, the magnetization (**M**) and elastic strain (**ε**) of the magnon resonator (e.g., YIG). The propagation of $\mathbf{H}^{EM}$ is governed by the Maxwell's equations, while dynamical evolution of $\mathbf{M}$ and $\boldsymbol{\varepsilon}$, which represent the magnon and phonon subsystem, are usually described by the Landau-Lifshitz-Gilbert (LLG) equation and elastodynamic equation, respectively. Crucially, the dynamics of $\mathbf{M}$ is modulated by the $\mathbf{H}^{EM}$ via the Zeeman torque, while the $\mathbf{M}$ and $\boldsymbol{\varepsilon}$ are coupled via the magnetoelastic interaction[30]. Therefore, a complete, direct numerical simulation of the dynamical processes in a cavity electromagnonic system requires the simultaneous solution of the coupled LLG, elastodynamic, and Maxwell's equations.

Thus far, there are only a few advanced computational models that include coupled dynamics of $\mathbf{M}$ and EM wave[31–34] but excludes either the exchange coupling field (i.e., macrospin approximation) in the LLG equation[31,33,34] or the displacement current in the Maxwell's equations[32]. Recently, models that include coupled dynamics of $\boldsymbol{\varepsilon}$, $\mathbf{M}$, and EM wave have also appeared[35–39], but these models are limited to 1D[36–38] or 2D[39] system or employ the Newton's equation[35] as a simplification of the elastodynamic equation. Furthermore, these models[35–39] have not yet been applied to a cavity electromagnonic system. The present dynamical phase-field model addresses the coupled dynamics of $\boldsymbol{\varepsilon}$, $\mathbf{M}$, and EM wave in a 3D cavity electromagnonic system by solving the coupled LLG, elastodynamic, and Maxwell's equations (see "Methods"). All numerical solvers are accelerated by graphics processing unit (GPU) to increase the computation throughput. As examples of application, we use the dynamical phase-field model to simulate the dynamics of excitation and control of magnon polariton modes in a cavity electromagnonic system comprised of a YIG magnon resonator placed in a 3D photon cavity. Typical coherent gate operations including Rabi oscillation and Ramsey interference are computationally demonstrated. Furthermore, we design a cavity electro-magno-mechanical system, which contains a bilayer YIG/SiN membrane placed in a 3D photon cavity and permits a resonant interaction between the magnon polaritons and the acoustic phonons. We then use the dynamical phase-field model to simulate the coupled mode dynamics under such triple phonon-magnon-photon resonance condition.

**Results**

*Simulation system set-up*

Figure 1a schematically shows the cavity electromagonic system. YIG, which has been widely used in hybrid magnonic systems[16–18,22,40–44] due to its ultralow magnetic damping, is used as the magnon resonator. The 3D microwave cavity has a dimension of 45×9×21 mm$^3$, which supports the TE$_{101}$ mode of the standing EM waves with a frequency $\omega_c/2\pi$=7.875 GHz. To excite the TE$_{101}$ cavity mode, a point charge current pulse $\mathbf{J}^c(t)$ in the form of a Gaussian function $te^{-t^2/2\sigma_0^2}$ is applied along the *y* axis (i.e., only the $J_y^c$ component is nonzero) at the position (22.5mm, 1.5mm, 10.5mm) of the cavity, where $\sigma_0$ is a free parameter that controls the pulse duration and chosen to be 70 ps so that the frequency window of the pulse covers the $\omega_c$. The simulated $\mathbf{H}^{EM}$ has a vortex-like distribution in the *xz* plane, as shown in Fig. 1a. The YIG resonator is placed at 1.5 mm below the top surface center of the cavity, where the magnitude of $\mathbf{H}^{EM}$ is relatively large. At the initial equilibrium state, the magnetization in the YIG $\mathbf{m}^0$ is along +*z* ([001]) due to a bias magnetic field $\mathbf{H}^{bias}$=(0,0,$H_z^{bias}$) applied along the same direction. Note that energy dissipation of both the cavity photon (arising from the imaginary component of the relative dielectric permittivity tensor $\varepsilon_r$) and magnon (arising from the effective magnetic damping) are both set to be zero to study the magnon-photon coupling under the most ideal situation.

The entire system is discretized into three-dimensional (3D) computational cells with a cell size $\Delta x=\Delta y=\Delta z=2$ nm. Numerically, a nm-scale computational cell is necessary to ensure spatial uniformity of the magnetization (i.e., the formation of Kittel mode magnon) via the $\mathbf{H}^{exch}$ (proportional to $|\nabla \mathbf{m}|^2$) between neighboring spins. Moreover, a basis for any micromagnetic simulation is that the cell size needs to be smaller than the exchange length $l_{ex}=\sqrt{A_{ex}/(0.5\mu_0 M_s^2)}$, which is about 16.3 nm for YIG with an exchange coupling coefficient $A_{ex}$ of 3.26 pJ/m and $M_s$= 140 kA/m[45]. However, discretization of a 45×9×21 mm$^3$ system using a cell size of $\Delta x=\Delta y=\Delta z=2$ nm would lead to a total of about $10^{21}$ cells, which is computationally unaffordable. To address the issue, the $\varepsilon_r$ is tuned to scale down the EM wavelength and hence the size of the microwave cavity. For example, in the case of 1×1×1 mm$^3$ YIG cube, we set all three diagonal components of the $\varepsilon_r$ to be 6.25×10$^{10}$ for both the YIG cube and the microwave cavity, thus the EM wavelength is scaled down by 2.5×10$^5$ (=$\sqrt{\varepsilon_r}$) times. Accordingly, the size of the microwave cavity can be reduced from 45×9×21 mm$^3$ to 180×36×84 nm$^3$ (i.e., 68,040 cells) without changing the spatial profile and the frequency of the TE$_{101}$ mode EM wave. Meanwhile, the size of the 1×1×1 mm$^3$ YIG cube should be scaled down to 4×4×4 nm$^3$ to maintain a constant volume ratio of the YIG cube to microwave cavity. Although such size down-scaling makes it not possible to simulate the high-order magnon modes (spatially non-uniform precession of local magnetization) that may occur in a mm-scale YIG, it would not influence the present work on the interaction between Kittel mode magnon and cavity photon. Importantly, although the larger $\varepsilon_r$ leads to a smaller $\mathbf{E}^{EM}$, the magnitude of $\mathbf{H}^{EM}$, which interacts with the magnetization, remains constant (see Eqs. (6-7) in "Methods"). As a result, the simulated coupled magnon-photon dynamics remains the same as that in the original mm-scale system. Furthermore, the use of a larger $\varepsilon_r$ allows using a larger time step which significantly reduces the computation time in long-term dynamics simulation (see "Methods").

*Magnon-photon coupling*

To demonstrate the validity and high numerical accuracy of our computational model, we first simulate the formation of the commonly observed $k$=0 mode magnon polariton ($k$ is wavenumber), which features the hybridization between the $k$=0 (Kittel) mode magnon $\hat{m}$ and $k\approx0$ mode cavity photon $\hat{c}$. As illustrated in Fig. 1a, the $\mathbf{H}^{EM}$, which is perpendicular to the initial equilibrium magnetization $\mathbf{m}^0$ (see Fig. 1a), drives the magnetization precession. Due to the exchange coupling, all local magnetization vectors $\mathbf{m}$ in the YIG process in phase, resulting in the excitation of the desirable Kittel mode magnon. Since $\mathbf{H}^{EM}$ is largely uniform around the YIG cube (*i.e.*, wavenumber $k\approx0$) and the magnon-photon interaction time is sufficiently long in the present 3D cavity, the $k$=0 mode magnon polariton should form if the angular frequency of the Kittel mode magnon $\omega_m$, or the ferromagnetic resonance (FMR) frequency, can be magnetically tuned to match the angular frequency of the cavity $\omega_c$. Specifically, for an initial equilibrium magnetization along [001], one has $\omega_m/2\pi=\gamma(H_z^{bias}-\frac{1}{3}M_s + \frac{2K_1}{\mu_0 M_s})$[36], where $\gamma$=27.86 GHz/T is the gyromagnetic ratio and $K_1$=620 J/m$^3$ is the magnetocrystalline anisotropy coefficient of the YIG. Accordingly, $\mathbf{H}^{bias}$=(0, 0, 0.291 T) is applied to have $\omega_m/2\pi=\omega_c/2\pi$=7.875 GHz.

Figure 1b shows the dynamics of the Kittel magnon mode and the photon, where $\Delta t$=0 refer to the moment at 20 ns after the injection of the Gaussian-shaped current pulse $\mathbf{J}^c(t)$ at $t$=0 ns. Typical behavior of coherent beating oscillation similar to a two-level system[37,46] is observed. Specifically, the peak amplitudes of the two modes (indicated by the trend lines) show a Rabi-like oscillation[17] with a period of ~6.9 ns (frequency ~ 145 MHz), suggesting a back-and-forth energy transfer between the YIG and the microwave cavity. Note that we focus on the peak amplitudes rather than the instantaneous values of $\Delta m_x$ and $E_y^{EM}$, because the energy of the magnon mode is swapped instantaneously between $\Delta m_x$ and $\Delta m_y$ while the energy of the cavity photon mode is swapped instantaneously between $E_y^{EM}$ and $H_x^{EM}$. For clearer illustration, Fig. 1c shows the simulated magnon state and spatial distribution of the radiation electric field $\mathbf{E}^{EM}(t)$ at a few representative moments. As shown in the left panel, when the $\mathbf{E}^{EM}$ reaches its peak amplitude, the magnetization $\mathbf{m}$ aligns almost along its initial direction [001], indicating an almost zero free energy change in the YIG. As the energy is being transferred from the cavity to the YIG, the amplitude of $\mathbf{E}^{EM}$ in the cavity decreases while the amplitude of the precessing magnetization (or $|\Delta\mathbf{m}|=\sqrt{(\Delta m_x)^2 + (\Delta m_y)^2}$ increases, as shown in the middle panel. After a half period of the energy transfer (3.45 ns = 6.9 ns/2), almost all the EM wave energy is absorbed by the YIG, which is indicated by the negligibly small $\mathbf{E}^{EM}$ in the cavity and relatively large $|\Delta\mathbf{m}|$, as shown in the right panel of Fig. 1c. Figure 1d shows the frequency spectrum of the temporal waveform $\Delta m_x(t)$ in Fig. 1b, which reveals two peak frequencies at 7.8 GHz and 7.945 GHz, respectively. The two peak frequencies are symmetric with respect to the $\omega_m/2\pi=\omega_c/2\pi=7.875$ GHz with a frequency gap of 145 MHz, indicating the formation of magnon polariton with two different hybrid modes $\hat{d}_+$ and $\hat{d}_-$. The frequency gap (denoted as $\delta_D$) is consistent with the frequency of the Rabi-like oscillation and defines a magnon-photon coupling strength $g_{cm}=\delta_D/2=2\pi\times72.5$ MHz. It is worth remarking that the magnon subsystem of the YIG is also coupled to the phonon subsystem, because the precessing $\mathbf{m}$ generates dynamical strain via the magnetoelastic feedback and the dynamical strain in turn modulates the dynamics of $\mathbf{m}$ via the $\mathbf{H}^{mel}$ (see Part 1 of the "Methods"). Moreover, the stiffness damping coefficient $\beta$ in the elastodynamic equation creates an additional channel for energy dissipation. However, in the system shown in Fig. 1a, the energy exchange between magnon and phonon subsystems is negligible because the magnitude of the dynamical strain is negligibly small (~$10^{-7}$) due to the relatively small magnetoelastic coupling constant of the YIG.

Figure 1e shows the numerically simulated mode frequencies (indicated by hollow circles) as functions of the bias magnetic field in three different hybrid systems where the sizes of the YIG cube are 0.4×0.4×0.4 mm³, 1×1×1 mm³ and 2×2×2 mm³, respectively, and the cavity size remains to be 45×9×21 mm³. The goal of these simulations is to computationally verify the theoretical relation of $g_{cm} = g_0\sqrt{N}$ (Ref. [16]), where $g_0$ is the coupling strength of a single Bohr magneton to the cavity; $N$ is the total spin number in the YIG and increases linearly with its size. The presence of avoided crossings in all three systems indicate the formation of magnon polaritons. The corresponding magnon-photon coupling strength $g_{cm}$ can be extracted from the frequency gap under on-resonance $\hat{m}$ and $\hat{c}$ modes (where $\Delta H^{bias}=0$), which are $2\pi\times18.3$ MHz, $2\pi\times72.5$ MHz, and $2\pi\times182.2$ MHz, respectively. One can evaluate that the $g_{cm}$ is largely proportional to the square root of the cube size and hence the $\sqrt{N}$. The $g_{cm}$ in the case of 2×2×2 mm³ is smaller than the value of $2\pi\times204.6$ MHz obtained from linear extrapolation due to the <100% spatial mode profile overlapping, which is consistent with experimental observation[17]. Based on the extracted $g_{cm}$, we fit the simulations results via the expression $\omega_\pm=\frac{1}{2}(\omega_m+\omega_c)\pm\frac{1}{2}\sqrt{(\omega_m-\omega_c)^2+4g_{cm}^2}$, which describe the angular frequencies of the two hybrid modes ($\hat{d}_+$ and $\hat{d}_-$) in a two-level system[47]. The excellent fitting indicates the validity of our model set-up and high numerical accuracy of our dynamical phase-field model.

*Floquet-induced magnonic Autler-Townes splitting*

Based on the cavity electromagnonic system with a YIG resonator of 1×1×1 mm$^3$, a periodic dynamical magnetic field $\mathbf{h}_D(t)$ is applied along the same axis (*z*) with the $\mathbf{H}^{bias}$ to implement Floquet engineering. The $\mathbf{h}_D(t)=|\mathbf{h}_D|\sin(\omega_D t)$ is applied uniformly on the YIG resonator, where $|\mathbf{h}_D|$ and $\omega_D$ are the amplitude and the angular frequency of the Floquet drive, respectively. Figure 2a shows the frequency spectra of the $\Delta m_x(t)$ simulated under a fixed $|\mathbf{h}_D|$ of 2000 A/m but different $\omega_D/2\pi$ varying from 0 to 400 MHz. A static bias magnetic field of 0.2315 MA/m (~0.291 T) is applied to have the magnon mode on resonance with the cavity photon and generate two magnon polariton modes $\hat{d}_+$ and $\hat{d}_-$, which have frequencies of 7.945 GHz and 7.8 GHz, respectively. Each polariton mode has several sidebands created by the Floquet drive. The frequencies of these sidebands either increase or decrease as the $\omega_D$ increases. When $\omega_D$ is equal to the (cavity and magnon) mode splitting frequency (=$2g_{cm}=2\pi\times145$ MHz), the first inner sideband of the magnon polariton mode $\hat{d}_+$ ($\hat{d}_-$) resonantly interact with the other magnon polariton mode $\hat{d}_-$ ($\hat{d}_+$). As a result, the two energy levels corresponding to the $\hat{d}_+$ and $\hat{d}_-$ modes split into four energy levels associated with different Floquet modes (as illustrated in the inset), where the frequency gap between the two newly split energy levels is denoted as $\Delta\omega_{AT}$, with $\Delta\omega_{AT}/2\pi=34.8$ MHz. Such magnonic Autler-Townes splitting[21] and the onset of avoided crossing indicate the realization of strong coupling between the two hybrid modes of the magnon polaritons by Floquet drive.

To gain further insights on the simulated spectrum, we analytically calculate the absorption spectrum of the magnon mode based on Floquet theory described in Ref. [21], using the numerically simulated on-resonance frequency $\omega_0$ (=$\omega_m=\omega_c$) and the magnon-photon coupling strength $g_{cm}$ as the inputs. As shown in Fig. 2b, the main structure of the calculated spectrum (shown in Fig. 2b) reproduces the simulation results. The calculated spectrum can be understood by writing the Floquet Hamiltonian as following (details of derivation are in Ref. [21]),

$$\frac{\hat{H}}{\hbar} = \omega_+ \hat{d}_+^{+}\hat{d}_+ + \omega_- \hat{d}_-^{+}\hat{d}_- + g_{cm}\sum_{n=odd} J_n(\frac{\Omega}{\omega_D})\left[\hat{d}_+^{+}\hat{d}_- e^{i(n\omega_D t)} + \hat{d}_+\hat{d}_-^{+} e^{-i(n\omega_D t)}\right]. \quad (1)$$

Here $\omega_\pm = \omega_0 \pm g_{cm}$, $\Omega = \gamma|\mathbf{h}_D|$, and $\hbar$ is the reduced Planck's constant. A list of symbols for various modes and related quantities is provided in Supplementary Note 1. The Floquet drive creates a series of sidebands of the $\hat{d}_\pm$ modes at frequencies $\omega_\pm \pm n\omega_D$, where *n* is the sideband order. The last term on the right-hand side of Eq. (1) describes the interactions between different sidebands, where $J_n$ is the *n*th Bessel function of the first kind. From Eq. (1), one can determine that the coupling strength between the first inner sideband of $\hat{d}_-$ and the $\hat{d}_+$ mode (i.e., the magnonic Autler-Townes splitting $\Delta\omega_{AT}$) is approximated as $\Delta\omega_{AT} \approx 2g_{cm}J_1(\Omega/\omega_D)$. Plugging in the numbers yields a theoretically predicted value of $\Delta\omega_{AT}/2\pi \sim 34.1$ MHz, which is in good agreement with the simulated value of 34.8 MHz. Moreover, since the sum in Eq. (1) only involves odd terms, some of the sidebands (*e.g.*, the first inner sideband of $\hat{d}_+$ and $\hat{d}_-$ modes) are not directly coupled, which is revealed by the crossing in both the simulated and calculated spectrum. Detailed discussion on this point can be found in Ref. [21].

As $|\mathbf{h}_D|$ increases, $\Delta\omega_{AT}$ varies in an oscillatory fashion but always remain nonzero (see Supplementary Fig. S1a in Supplementary Note 2). This trend cannot be explained by the analytical approximation $\Delta\omega_{AT} \approx 2g_{cm}J_1(\Omega/\omega_D)$, which is only valid when $|\mathbf{h}_D|$ is relatively small. At large $|\mathbf{h}_D|$, the $|\mathbf{h}_D|$-dependent $\Delta\omega_{AT}$ can be better quantified by the following analytical expression (see detailed derivation in Supplementary Note 3),

$$\Delta\omega_{AT} = 2g_{cm}\left(1 - J_0\left(\frac{\Omega}{\omega_D}\right)\right), \qquad (2)$$

which predicts a repeated occurrence of $\Delta\omega_{AT} = 2g_{cm}$ at $J_0 = 0$ and the presence of a Floquet ultrastrong coupling regime[21] where $\Delta\omega_{AT} > 2g_{cm}$ and $J_0 < 0$. Both features are shown in the frequency spectrum of the magnon polaritons (Supplementary Fig. S1a) obtained by dynamical phase-field simulations.

Our dynamical phase-field simulation results in Fig. 2c further shows the temporal profile of the $\Delta m_x(t)$ and $E_y^{EM}(t)$ for $\omega_D/2\pi = 300$ MHz. According to the frequency spectra in Fig. 2a, the magnon polariton is still dominated by the intrinsic hybrid modes $\hat{d}_+$ and $\hat{d}_-$ with no magnonic Autler-Townes splitting. Correspondingly, the amplitudes of both the $\Delta m_x(t)$ and $E_y^{EM}(t)$ display a Rabi-like oscillation with a period of 6.9 ns, which is the same as in Fig. 1b. By comparison, for $\omega_D/2\pi=145$ MHz where the magnonic Autler-Townes splitting is prominent, the corresponding temporal profiles of $\Delta m_x(t)$ and $E_y^{EM}(t)$, as shown in Fig. 2d, are clearly composed of components of more than two frequencies. Specifically, there are four major frequency components at $\omega_+ + 0.5\Delta\omega_{AT}$, $\omega_+ - 0.5\Delta\omega_{AT}$, $\omega_- + 0.5\Delta\omega_{AT}$, and $\omega_- - 0.5\Delta\omega_{AT}$, respectively, corresponding to the four split energy levels as shown in Fig. 2a. Despite the more complex temporal profile, the evolution of the peak amplitudes of $\Delta m_x(t)$ and $E_y^{EM}(t)$ are still complementary, indicating that the back-and-forth energy exchange still occurs between the Kittel magnon mode $\hat{m}$ and the cavity photon mode $\hat{c}$. The beam-splitter type coupling between the $\hat{d}_+$ ($\hat{d}_-$) and the first inner sideband of the $\hat{d}_-$ ($\hat{d}_+$) mode (Fig. 2a) can also be interpreted as the energy exchange (i.e., Rabi-like oscillation) between the two energy levels (i.e., the hybridized modes $\hat{d}_+$ and $\hat{d}_-$) through the frequency matching provided by the Floquet drive, i.e., $\omega_+ = \omega_- + \omega_D$. Similar Floquet-driven Rabi-like oscillation between two hybridized modes has also been demonstrated experimentally in a two-level photonic system[27].

*Dynamical control of the energy exchange rate between two magnon polariton modes*

The Rabi-like oscillation between the two magnon polariton modes ($\hat{d}_+$ and $\hat{d}_-$) corresponds to a rotation along the real axis of a Bloch sphere, as illustrated in Fig. 3a inset. Here, we model this process in the time domain and computationally demonstrate the dynamical control of the energy exchange rate between the $\hat{d}_+$ and $\hat{d}_-$ modes (namely, Rabi flopping frequency) by dynamically varying the amplitude $|\mathbf{h}_D|$ of the Floque drive. To this end, the system is initialized by pumping a 15-cycle sinusoidal charge current pulse $\mathbf{J}^c(t)=J_0\sin(\omega t)$ at $\omega=2\pi\times 7.945$ GHz along the y axis (i.e., only the $J_y^c$ component is nonzero) to populate (excite) the $\hat{d}_+$ mode. A continuous Floquet drive $\mathbf{h}_D(t)$ at the cavity-magnon mode splitting frequency ($\omega_D/2\pi=\delta_D/2\pi=145$ MHz) is then applied to the system. Figure 3a shows the temporal evolution of the magnetization amplitude of the $\hat{d}_+$ mode magnon polariton (denoted as $|\Delta\mathbf{m}|_{\hat{d}_+}$) with $|\mathbf{h}_D|=5000$ A/m. Here $|\Delta\mathbf{m}|_{\hat{d}_+}(t)$ is obtained by first extracting the temporal evolution of $\Delta m_x(t)$ and $\Delta m_y(t)$ by performing inverse Fourier transform for 7.945-GHz (±50 MHz) peak in their frequency spectra and then calculating its magnetization amplitude via $\sqrt{(\Delta m_x)^2 + (\Delta m_y)^2}$. The $|\Delta\mathbf{m}|_{\hat{d}_+}(t)$ displays a Rabi-like oscillation with a period of 11.6 ns, corresponding to a flopping frequency of 86.2 MHz. We note

that the $\hat{d}_-$ mode magnon polariton (at 7.8 GHz) was also excited after the initial current pulse injection, and the dynamics of $|\Delta \mathbf{m}|_{\hat{d}_-}(t)$ complements the $|\Delta \mathbf{m}|_{\hat{d}_+}(t)$, as shown by Supplementary Fig. S2 in Supplementary Note 4, suggesting a dynamical energy exchange between the $\hat{d}_+$ and $\hat{d}_-$ modes.

Figure 3b further summarizes the $|\Delta \mathbf{m}|_{\hat{d}_+}(t)$ simulated under different $|\mathbf{h}_D|$ varying from 1000 to 5000 A/m. As shown, the period of Rabi-like oscillation decreases as $|\mathbf{h}_D|$ increases, leading to an increase in the corresponding Rabi flopping frequency ($f_{Rabi}$). By analytically solving the Heisenberg equation of the $\hat{d}_+$ and $\hat{d}_-$ modes under the rotating wave approximation (RWA), we obtain the mode square $|\hat{d}_+(t)|^2 = \cos^2\left(\frac{\Omega}{4}t\right)$ (see Supplementary Note 5), yielding $f_{Rabi}=\Omega/2=\gamma|\mathbf{h}_D|/2$, which is only valid when $\omega_D=2g_{cm}=2\pi\times 145$ MHz. It is worth noting that Figure 3a shows a nonzero offset, which differs from the RWA-based prediction and may be due to the existence of other magnon modes. As shown in the inset of Fig. 3b, the analytically calculated $f_{Rabi}$ agrees well with the values extracted from the simulated $|\Delta \mathbf{m}|_{\hat{d}_+}(t)$ with only small deviation at larger $|\mathbf{h}_D|$ values, where the system's behavior deviates from the RWA.

As $|\mathbf{h}_D|$ further increases, it is no longer possible to numerically extract the $|\Delta \mathbf{m}|_{\hat{d}_+}(t)$ and hence the $f_{Rabi}$ because multiple magnon modes coexist at 7.945 GHz. However, it is expected that that the variation trend of $f_{Rabi}$ with $|\mathbf{h}_D|$ would deviates significantly from the RWA-predicted linear relation. Specifically, it is reasonable to speculate that the $f_{Rabi}\approx \Delta\omega_{AT}/2\pi$ because the present Rabi oscillation is based on beam-splitter type coupling between the $\hat{d}_+$ ($\hat{d}_-$) and the first inner sideband of the $\hat{d}_-$ ($\hat{d}_+$) mode, as discussed above. In this regard, the $|\mathbf{h}_D|$-dependent $f_{Rabi}$ should follow the $|\mathbf{h}_D|$-dependent $\Delta\omega_{AT}$, which is oscillatory at large $|\mathbf{h}_D|$ as shown in Supplementary Fig. S1a.

*Dynamical control of the relative phase between two magnon polariton modes*

Experimentally, it has also been shown that driving the transition between two hybridized modes of a two-level photonic system with detuned pulses enables a dynamical control over the relative phase of the two hybrid modes[27]. By analogy to the protocols described in Ref. [27], here we computationally demonstrate the dynamical control of the relative phase between $\hat{d}_+$ and $\hat{d}_-$ mode magnon polariton modes (namely, magnonic Ramsey interference). We first excite the system to the $\hat{d}_+$ mode by injecting a 15-cycle sinusoidal $\mathbf{J}^c(t)$ at 7.945 GHz. A $\pi/2$ pulse of the Floquet drive field $\mathbf{h}_D(t)$ with a frequency of $\omega_D=\delta_D+\Delta\omega$ is then applied to create excitations in a superposition of both the $\hat{d}_+$ and $\hat{d}_-$ modes, which is illustrated by State 'I' on the equator of the Bloch sphere (see Fig. 4a). Here, the duration of the $\pi/2$ pulse is 1/4th of the Rabi-like oscillation period under the field amplitude $|\mathbf{h}_D|$=5000 A/m, i.e., $\tau_0$=1/(4$f_{Rabi}$)=2.9 ns. After the first $\pi/2$ pulse is turned off ($\mathbf{h}_D(t) = 0$), the excitation as superposition of $\hat{d}_+$ and $\hat{d}_-$ modes would start to precess along the equator of the Bloch sphere (i.e., free evolution) with a precession frequency determined by the detuning amplitude $\Delta\omega$. Upon the completion of the free evolution period, a second $\pi/2$ pulse is applied to project the excitations (State 'II' in the Bloch sphere) into the $\hat{d}_-$ mode. The amplitude of the $\hat{d}_-$ mode after the completion of second $\pi/2$ pulse is determined by the relative phase between the two modes. The relative phase difference is proportional to both the frequency detuning $\Delta\omega$ and the duration of free evolution $\tau$. To computationally demonstrate this principle, we record the magnetization amplitude of the $\hat{d}_-$ mode ($|\Delta \mathbf{m}|_{\hat{d}_-}$) after the completion of the second $\pi/2$ pulse, as a function of the free evolution duration $\tau$ under $\Delta\omega/2\pi$ =10,15 and 20 MHz. As shown in Fig. 4b, the frequency for the variation of the $|\Delta \mathbf{m}|_{\hat{d}_-}$ with the duration $\tau$ is exactly equal to $\Delta\omega$. The result matches the analytical expression, $\left|\hat{d}_-(t)\right|^2 = \frac{|\alpha|^2}{2}\sin^2\left(\frac{\Delta\omega}{2}\tau\right)$ (see Supplementary Note 6), where $\alpha$ is the initial mode amplitude in $\hat{d}_+$.

*Coupled mode dynamics under the triple phonon-magnon-photon resonance condition*

To simulate the coupled mode dynamics under the triple resonance condition, we design an electro-magno-mechanical system which contains a YIG/SiN bilayer membrane placed in a 3D photon cavity, as shown in Fig. 5a. The cavity hosts a nominal $TE_{100}$ mode photon (i.e., the EM wave has a half-wavelength profile along the $x$ axis while is spatially uniform along $y$ and $z$), which is a smaller portion of a larger-scale $TE_{101}$ mode cavity. Details of the system design and simulation set-up are shown in the Methods section. The resonant frequency of the cavity photon (~ 9.1 GHz) is the same as the frequency of the Kittel mode magnon to enable the formation of magnon polaritons. In the present set-up, the frequencies of the two magnon polariton modes are found to be 9.028 GHz ($\hat{d}_-$) and 9.14 GHz ($\hat{d}_+$), respectively. Based on the magnetoelastic backaction, the dynamically processing magnetization of the Kittel mode magnon in the YIG layer will generate chiral transverse acoustic (TA) phonons that has a wavevector along the thickness direction ($y$) of the bilayer, as has been demonstrated experimentally in a similar YIG/GGG bilayer[22,48]. To obtain a large profile overlap between phonons and magnons, the layer thicknesses of the YIG (10 nm)/SiN (270 nm) bilayer membrane are designed to host a fundamental ($n=1$) TA phonon of 9.13 GHz. Because this frequency is close enough to the frequency of the $\hat{d}_+$ mode magnon polariton, resonant interaction between the chiral TA phonon and the $\hat{d}_+$ mode magnon polariton can be enabled. Such triple resonance among the fundamental TA phonon, the Kittel mode magnon, and the $k≈0$ mode photon is similar to the experiment by Zhang *et al.*[15] where the TA phonon of a much higher order ($n=2960$) interacts with the Kittel mode magnon, resulting in a smaller mode profile overlap and hence a lower magnon-phonon coupling strength than the present design.

The vectors in Fig. 5a show the direction and the magnitude of the local $\mathbf{H}^{EM}$ in the cavity. The hybridization of the Kittel mode magnons and cavity photons alters the local EM fields in the vicinity of the YIG resonator, as shown more clearly in Fig. 5b. Figure 5c shows the spatial profile of the fundamental TA phonon across the YIG/SiN bilayer. The right-handed phonon chirality is shown in Fig. 5d. Figure 5e,g,i show the evolution of the $H_z^{EM}$ in the cavity, the $\Delta m_x$ in the YIG, and the local $\varepsilon_{xy}$ in the SiN, and their frequency spectra are shown in Fig. 5f,h,j, respectively. As shown, the precessing magnetization of both the $\hat{d}_-$ and $\hat{d}_+$ mode magnon polaritons will generate chiral TA phonons of the same frequencies at 9.028 GHz and 9.14 GHz, respectively. The populations of the $\hat{d}_-$ and $\hat{d}_+$ modes are similar, as indicated by the similar spectral amplitudes of these two peaks in the frequency spectra of photons and magnons (see Figs. 5f,h). However, the population of the 9.14 GHz phonon, due to its proximity to the intrinsic phonon resonance frequency of 9.13 GHz, is significantly larger than that of the 9.028 GHz phonon (see Fig. 5j). Interestingly, there exists a strong peak at 9.13 GHz in the phonon frequency spectrum, even though the population of the driving $\hat{d}_+$ mode magnon polariton is low at 9.13 GHz. More interestingly, there exist small peaks at 9.13 GHz in the frequency spectra of both the $H_z^{EM}$ and the $\Delta m_x$, indicating an energy backflow from the magnetically excited 9.13 GHz phonon mode to both the magnon and photon systems. This energy backflow, which becomes pronounced in this case mainly because the damping terms for all three systems are turned off (i.e., no energy dissipation), is clear evidence of the triple phonon-magnon-photon resonance. As a control simulation (see

Supplementary Fig. S3 in Supplementary Note 7), we found that turning on the elastic damping for the YIG and SiN leads to a significantly lower population for the magnetically existed 9.13 GHz phonon mode, and that there are no additional peaks at 9.13 GHz in the spectra of magnon and photons.

**Discussion**

We have developed a 3D dynamical phase-field model that incorporates the coupled dynamics of strain, magnetization, and EM wave in a cavity electromagnonic system, which integrate magnon/phonon resonator(s) in a bulk 3D photon cavity. By solving the coupled equations of motion for these quantities under appropriate magnetic, mechanical, and EM boundary conditions, our computational model allows predicting the spatiotemporal evolution of strain, magnetization, and EM fields under various operating conditions directly from the fundamental material parameters. As examples, time-domain dynamics of relevant modes in typical coherent gate operations (Rabi oscillation and Ramsey interference) are simulated. The physical validity and high numerical accuracy of the solvers for coupled magnon-photon dynamics in our dynamical phase-field model were demonstrated by understanding the simulation results with analytical fitting and rigorous Hamiltonian-based Floquet theory. We have also applied the dynamical phase-field model to design a cavity electro-magno-mechanical system that enables the triple phonon-magnon-photon resonance, and computationally demonstrate the resonant excitation of a chiral, fundamental ($n$=1) TA phonon mode by magnon polaritons under such triple resonance condition.

In combination with the high throughput resulting from the GPU acceleration, the present 3D dynamical phase-field model can be used to guide the experimental design of the microwave photon cavity, magnon and phonon resonator(s) as well as the operating condition for the discovery of new physical phenomena as well as the optimization of key device features such as the coupling strength, mode swapping rate (e.g., Rabi flopping frequency), and cooperativity. The present dynamic phase-field model can also be extended to design and simulate cavity electromagnonic, magnomechanical, and electro-magno-mechanical systems with more complex structures, such as the on-chip systems integrating a coplanar microwave resonator and a magnon resonator[11–14], by implementing a more detailed treatment of the current dynamics in the normal/superconducting metal components (e.g., see a relevant recent modeling work[49]).

**Methods**

*Part 1: Description of the 3D dynamical phase-field model incorporating coupled dynamics of strain, magnetization, and EM waves in multiphase systems.*

A phase-field model leverages the symmetry-consistent use of continuum physical order parameters and their gradients to describe the total free energy of a spatially inhomogeneous system. The functional derivative of the total free energy ($F_{\text{tot}}$) with respect to a specific order parameter yields the thermodynamic driving force that drives the evolution of the order parameter. For example, the effective magnetic field that drives the evolution of **M** is calculated as $\mathbf{H}^{\text{eff}} = -\frac{1}{\mu_0}\frac{\delta F_{\text{tot}}}{\delta \mathbf{M}}$, where $\mu_0$ is vacuum permeability. In a dynamical phase-field model, equations of motion for all key order parameters are typically solved in their exact forms[36,37,50]. This is different from conventional phase-field model, where faster-evolving order parameters are often assumed to reach steady or equilibrium state instantaneously[51]. A dynamical phase-field model is particularly necessary for hybrid systems featuring bidirectional dynamical energy exchange and conversion between different physical subsystems, as in cavity electromagnonic systems.

As an example, we consider a commonly used system that contains a bulk magnon resonator in a 3D microwave cavity. The evolution of the normalized magnetization **m**= **M**/$M_s$ in the magnon resonator, where $M_s$ is the saturation magnetization, is governed by the LLG equation, i.e.,

$$\frac{\partial \mathbf{m}}{\partial t} = -\frac{\gamma}{1+\alpha_{\text{eff}}^2} \mathbf{m} \times \mathbf{H}^{\text{eff}} - \frac{\gamma \alpha_{\text{eff}}}{1+\alpha_{\text{eff}}^2} \mathbf{m} \times (\mathbf{m} \times \mathbf{H}^{\text{eff}}), \tag{3}$$

where $\gamma$ is the gyromagnetic ratio; $\alpha_{\text{eff}}$ is the effective magnetic damping coefficient; $\mathbf{H}^{\text{eff}}=\mathbf{H}^{\text{app}}+\mathbf{H}^{\text{anis}}+\mathbf{H}^{\text{exch}}+\mathbf{H}^{\text{mel}}+\mathbf{H}^{\text{d}}+\mathbf{H}^{\text{EM}}$ is the total effective magnetic field, where the externally applied magnetic field $\mathbf{H}^{\text{app}}(t)$ includes both the static bias magnetic field $\mathbf{H}^{\text{bias}}$ and a dynamic Floquet driving magnetic field $\mathbf{h}_D(t)$; $\mathbf{H}^{\text{anis}}$ and $\mathbf{H}^{\text{exch}}$ are the effective magnetic field resulting from the functional derivatives of the magnetocrystalline anisotropy energy and exchange coupling energy with respect to **m**, respectively, and their expressions are provided in our previous work[36]; $\mathbf{H}^{\text{mel}}$ is the effective magnetoelastic field, $\mathbf{H}^{\text{d}}$ is the demagnetization field, and $\mathbf{H}^{\text{EM}}$ is the magnetic field of the cavity electromagnetic wave. $\mathbf{H}^{\text{mel}}$ is calculated as $\mathbf{H}^{\text{mel}} = -\frac{1}{\mu_0}\frac{\partial f_{\text{elas}}}{\partial \mathbf{M}}$. Here the elastic free energy density $f_{\text{elas}} = \frac{1}{2} c_{ijkl}(\varepsilon_{kl} - \varepsilon_{kl}^0)(\varepsilon_{ij} - \varepsilon_{ij}^0)$, with $i, j = x, y, z$. Here the $c_{ijkl}$ is the elastic stiffness tensor; for magnets of cubic symmetry, the stress-free strain $\varepsilon_{ii}^0 = \frac{3}{2}\lambda_{100}\left(m_i^2 - \frac{1}{3}\right)$ and $\varepsilon_{ij}^0 = \frac{3}{2}\lambda_{111} m_i m_j$, where $\lambda_{100}$ and $\lambda_{111}$ are magnetostriction coefficients. The local total strain $\varepsilon_{ij}$ can be written as $\varepsilon_{ij}(t) = \varepsilon_{ij}^{\text{eq}} + \Delta\varepsilon_{ij}(t)$, where the $\varepsilon_{ij}^{\text{eq}}$ is the total strain at the initial equilibrium state and can be obtained by solving the mechanical equilibrium equation $\nabla \cdot \sigma_{ij}^{\text{eq}} = \nabla \cdot [c_{ijkl}(\varepsilon_{ij}^{\text{eq}} - \varepsilon_{ij}^{0,\text{eq}})] = 0$. For a stress-free, uniformly magnetized magnetic material, $\varepsilon_{ij}^{\text{eq}} = \varepsilon_{ij}^{0,\text{eq}}$. The dynamical strain $\Delta\varepsilon_{ij}$ is calculated via $\Delta\varepsilon_{ij} = \frac{1}{2}\left(\frac{\partial \Delta u_i}{\partial j} + \frac{\partial \Delta u_j}{\partial i}\right)$, and the time-varying local mechanical displacement $\Delta \mathbf{u} = \mathbf{u} - \mathbf{u}^{\text{eq}}$ is obtained by solving the elastodynamic equation,

$$\rho \frac{\partial^2 \Delta \mathbf{u}}{\partial t^2} = \nabla \cdot (\Delta \boldsymbol{\sigma} + \beta \frac{\partial \Delta \boldsymbol{\sigma}}{\partial t}), \tag{4}$$

where $\Delta\boldsymbol{\sigma}=\boldsymbol{\sigma}-\boldsymbol{\sigma}^{\text{eq}}$ is the dynamical stress; $\rho$ is the mass density, and $\beta$ is the stiffness damping coefficient. Since Eq. (4) is solved for the entire system, the material parameters $c_{ijkl}$, $\rho$, and $\beta$ vary in different phases, where the continuity boundary condition for **u** and $\boldsymbol{\sigma}$[52] are applied at the interface between the magnon resonator and the microwave cavity. In this regard, by setting the $c_{ijkl}$ of the microwave cavity to be zero, the stress-free surface of the magnon resonator is automatically considered. The magnon resonator is cube-shaped, which is a computationally more tractable geometry because the entire simulation system is discretized by cube-shaped cells. Based on the coupling to the LLG equation solver, this numerical solver of the elastodynamic equation has previously been applied to simulate coupled magnon-phonon dynamics both the 1D[36,37] and 2D [39] magnetic multilayer system. The high numerical accuracy of this elastodynamic solver have been demonstrated through comparison to the analytical solutions in 1D system or the 2D simulations performed via the commercial COMSOL Multiphysics® software. Here, we demonstrate the numerical accuracy of this elastodynamic solver in a 3D elastically inhomogeneous multiphase system through the comparison to the simulation results obtained from the COMSOL Multiphysics® (Supplementary Note 8).

The demagnetization (stray) field $\mathbf{H}^d$ can be expressed as $H_i^d(t) = H_i^{d,eq} + \Delta H_i^d(t)$. The $H_i^{d,eq}$ is produced by the magnetization $\mathbf{m}^0 = \mathbf{m}(t=0)$ at the initial equilibrium state inside the magnon resonator, and can be obtained by solving the continuity equation for magnetic flux $\nabla \cdot \mathbf{B}^{eq} = \nabla \cdot [\mu_0(\mathbf{H}^{d,eq} + \mathbf{m}^0 M_s)] = 0$, which is part of the Maxwell's equations. The magnetic boundary condition $\partial \mathbf{m}/\partial \mathbf{n} = 0$ is applied on the surfaces of the magnon resonator, where $\mathbf{n}$ is normal vector to the surface. For a cubic or spheric magnet with spatially uniform magnetization, $\mathbf{H}^{d,eq} = -\frac{1}{3} M_s(m_x^0, m_y^0, m_z^0)$. The dynamically changing $\Delta H_i^d(t)$, which emerges when $\mathbf{m}$ starts to evolve, does not need to be calculated separately. Rather, the magnetic-field component of the EM wave $\mathbf{H}^{EM}(t)$, which is obtained by solving the two dynamical equations in the Maxwell's equations via the finite-difference time-domain (FDTD) solver on Yee grid (to be discussed below), can automatically satisfy the magnetic flux continuity equation $\nabla \cdot \mathbf{B} = 0$ across the heterointerfaces[53]. Specifically, the spatiotemporal evolution of the $\mathbf{H}^{EM}$ and the associated electric field component of the EM wave $\mathbf{E}^{EM}$ are simulated by solving the Maxwell's equations,

$$\nabla \times \mathbf{H}^{EM} = \varepsilon_0 \varepsilon_r \frac{\partial \mathbf{E}^{EM}}{\partial t} + \mathbf{J}^c , \quad (5)$$

$$\nabla \times \mathbf{E}^{EM} = -\frac{\partial \mathbf{B}}{\partial t} = -\mu_0 \left( \frac{\partial \mathbf{H}^{EM}}{\partial t} + M_s \frac{\partial \mathbf{m}}{\partial t} \right). \quad (6)$$

Equation (5-6) indicates that the $\mathbf{H}^{EM}$ is produced by both the free charge current pulse $\mathbf{J}^c(t)$ via electric dipole radiation and the precessing $\mathbf{m}(t)$ via magnetic dipole radiation. The perfect electric conductor (PEC) boundary condition is applied on all surfaces of the microwave cavity for reflecting the EM wave without loss. Specifically, $E_i = 0$ and $E_j = 0$ on the $ij$ surfaces of the cavity for PEC, with $i = x, y, z,$ and $j \neq i$. The elastic stiffness coefficients, the damping coefficient, and the mass density of YIG are listed in Supplementary Note 8. Other material parameters of YIG used in the simulations, including the magnetocrystalline anisotropy and the magnetoelastic coupling coefficients of YIG can be found in ref. [37]. Central finite difference is used for calculating spatial derivatives with a midpoint derivative approximation. Conventional Yee grid and the 3D FDTD method[53] are used to numerically discretize the EM wave and solve Eqs. (5-6). All dynamical equations are solved in a coupled fashion using the classical Runge-Kutta method with a time step $\Delta t = 5 \times 10^{-14}$ s. The choice of $\Delta t$ is subjected to the Courant condition for numerical convergence in conventional FDTD algorithm, which requires $\Delta t \leq l_0/(\sqrt{3}v)$[33], where $l_0$ is the simulation cell size and $v$ is the EM wave velocity in the medium. Since the use of a larger $\varepsilon_r$ leads to $\sqrt{\varepsilon_r}$ times smaller $v$ compared to speed of light in vacuum, a larger $\Delta t$ can be used in this work, which significantly reduces the computational time in long-term dynamics simulation.

*Part 2: Detailed discussion of the size scaling method*

Under the same excitation charge current, a larger $\varepsilon_r$ would also lead to an $\mathbf{E}^{EM}$ that is $\sqrt{\varepsilon_r}$ times smaller, because the amplitude of $\mathbf{E}^{EM}$ is inversely proportional to the angular wavenumber $k$ of the EM wave, with $k = \omega\sqrt{\varepsilon_0 \varepsilon_r \mu_0} = \omega/v$. This relationship between $\mathbf{E}^{EM}$ and $k$ can be quantitatively understood by rewriting $\mathbf{J}^c(t) = \partial \mathbf{P}/\partial t$ and then analytically solving the wave equation of the $\mathbf{E}^{EM}$ under the plane-wave assumption (see details in [54]).

Despite the smaller $\mathbf{E}^{EM}$, the amplitude of $\mathbf{H}^{EM}$ would remain unchanged because the ratio of $\mathbf{E}^{EM}$ and the $\mathbf{B}$ field is also related by the EM wave velocity $v$. For example, let us we focus on the $B_x$ ($H_x^{EM}$) component, which is the main component that interacts with the magnon mode in the system shown in Fig. 1a. We also consider that only the dominant $E_y^{EM}$ component is nonzero, which is consistent with Fig. 1c. In this case, Eq. (6) can be rewritten as,

$$\frac{\partial B_x}{\partial t} = -\left(\frac{\partial E_z^{EM}}{\partial y} - \frac{\partial E_y^{EM}}{\partial z}\right) \approx \frac{\partial E_y^{EM}}{\partial z} \qquad (7)$$

If we further write $B_x = B_x^0 e^{i(\omega t - kz)}$ and $E_y^{EM} = E_y^{EM,0} e^{i(\omega t - kz)}$ under the plane-wave assumption, Eq. (7) can be further rewritten into,

$$B_x = -E_y^{EM}\frac{k}{\omega} = -E_y^{EM}\sqrt{\varepsilon_0 \varepsilon_r \mu_0}. \qquad (8)$$

Therefore, although a larger $\varepsilon_r$ reduces the $E_y^{EM}$ by $\sqrt{\varepsilon_r}$ times, as shown in Eq. (8), the multiplication by $\sqrt{\varepsilon_r}$ leaves the $B_x$ unchanged. The $H_x^{EM}$ also remains unchanged, because (i) $B_x = \mu_0 H_x^{EM}$ in the cavity and $B_x = \mu_0(1 + \chi_m)H_x^{EM}$ in the YIG resonator ($\mathbf{M}=\boldsymbol{\chi}_m \mathbf{H}^{EM}$); and (ii) the magnetic susceptibility tensor $\boldsymbol{\chi}_m$ is independent of the $\varepsilon_r$.

To demonstrate the applicability of the conclusion above to cases that are more general than plane-wave assumption, we also scale down the size of the cavity electromagnonic system shown in Fig. 1a from the original size of 45×9×21 mm³ to a few different sizes, in addition to the 180×36×84 nm³ used in the main paper (for which $\varepsilon_r$ was increased from 1 to 6.25×10¹⁰). These additional sizes include 225×45×105 nm³, 300×60×140 nm³, and 450×90×210 nm³, where the $\varepsilon_r$ was increased from 1 to 4×10¹⁰, 2.25×10¹⁰, and 1×10¹⁰, respectively. We then perform dynamical phase-field simulations to model the excitation of magnon polaritons in these three systems in a similar manner to those in Fig. 1. As expected, the amplitude of $E_y^{EM}$ is $\sqrt{\varepsilon_r}$ times smaller while the $H_x^{EM}$ remains unchanged with the increasing $\varepsilon_r$, as shown by Supplementary Fig. S5 in Supplementary Note 9. By proportionally scaling down the size of the magnon resonator, the magnon-photon coupling strength would remain unchanged.

The main reason for scaling down a mm-scale system to the nm scale is to ensure that the magnon resonator only accommodates the Kittel mode magnon due to the dominant Heisenberg exchange coupling at the nm-scale. An alternative approach is to use one single cell to represent the magnon resonator (i.e., the macrospin approximation, as in Ref. [55]). We adopt this approach in the design of cavity electro-magno-mechanical system with triple phonon-magnon-photon resonance, as will be discussed below. In this approach, the Heisenberg exchange coupling does not need to be considered, and the spatial discretization of the system is mainly determined by the need to discretize an EM wave. This approach allows for simulating a large-scale photon cavity with a relatively small number of cells and for designing the size and shape of the cavity, but does not permit studying the effect of the size (e.g., as in Fig. 1e) and the shape of the magnon resonator on the coupled magnon-photon dynamics. If using multiple simulation cells yet turning off the exchange coupling, then there would be no force to lock the discrete local magnetization vectors into one giant spin. As time goes, what we have found for the present cavity electromagnonic system (Fig. 1a) is that the numerical error will accumulate (i.e., the values of $m_i$ will become more and more spatially nonuniform) and eventually lead to numerical divergence. Applying a stronger bias magnetic field would only alleviate this numerical issue by delaying, rather than preventing, the numerical divergence, and would also impose a constraint on the frequency range of the magnons that can be excited. This issue of numerical instability is even worse when the size of the magnon resonator is big enough such that the cavity magnetic field inside the resonator becomes highly inhomogeneous.

*Part 3: Design and simulation set-up of a 3D cavity electro-magno-mechanical system enabling triple phonon-magnon-photon resonance*

To excite a nominal $TE_{100}$ photon mode, we consider a cavity that is 16.5-mm-long (the half-wavelength of the 9.1 GHz EM wave at $\varepsilon_r=1$) in the *x* axis and apply periodic boundary conditions to the *xy* surfaces as well as the PEC boundary condition to all the other surfaces. This set-up is equivalent to placing an array of YIG/SiN bilayer membranes in a large cavity and then probing the phonon-magnon-photon coupling in one of the repeating units (see Supplementary Fig. S6a in Supplementary Note 10 for the full-scale cavity design). To minimize the interaction between the neighboring YIG resonators, we chose a length of 5.5 mm along the *z* axis, which is sufficiently long to ensure that the EM field remains largely uniform in the *xz* plane except the region near the YIG, as shown by the spatial distribution of the $H_z^{EM}$ in Fig. 5b. The simulated photon profile in Fig. 5a can be considered as a smaller portion of the profile of the $TE_{101}$ mode photon in a larger-scale cavity (see Supplementary Fig. S6b in Supplementary Note 10). Along the *y* axis of the cavity, which is parallel to the wavevector of the acoustic phonons and the thickness direction of the YIG/SiN bilayer, we consider a length of 400 nm to ensure the formation of the fundamental ($n=1$) TA phonon mode at the GHz frequency. The EM field is always uniform along *y* because the EM wavelength (33 mm) is far larger than the 400 nm. Taken together, the dimensions of the 3D cavity are 16.5 mm (*x*) × 400 nm (*y*) × 5.5 mm (*z*). The size of the YIG/SiN bilayer membrane is 0.55 mm (*x*) × 280 nm (*y*) × 0.5 mm (*z*), where the thicknesses of the YIG and SiN layer are 10 nm and 270 nm, respectively.

Simulating the GHz phonon dynamics along the *y* axis requires that the cell size is nm-scale along the *y* axis. According to the Courant condition for the FDTD algorithm mentioned above, a time step $\Delta t$ on the order of $10^{-18}$ s would be required to maintain numerical convergency. It would be computationally prohibitive to simulate the GHz mode dynamics over a time frame over hundreds of ns using such a small $\Delta t$. To address this issue, we increase the $\varepsilon_r$ from 1 to $10^8$, which allows for using a $10^4$ ($=\sqrt{\varepsilon_r}$) times larger $\Delta t$ due to the slower EM wave velocity as discussed earlier. In the meantime, the EM wavelength and hence the size of the photon cavity are reduced by $10^4$ ($=\sqrt{\varepsilon_r}$) times. Given that the original cavity is only 400-nm-long along the *y* axis, this would result in an impractical size of 0.04 nm. Alternatively, we consider (i) a 3D photon cavity of 1650 nm (*x*) × 400 nm (*y*) × 550 nm (z) with $\varepsilon_r=10^8$, where the dimensions of the cavity along the *x* and *z* axis were reduced by $10^4$ times yet the dimension along the *y* axis is kept the same as the original one; and (ii) a YIG/SiN bilayer of 55 nm (*x*) × 280 nm (*y*) × 50 nm (*z*) in the simulations, where the in-plane dimensions of the bilayer were reduced by $10^4$ times yet the thicknesses of the YIG (10 nm) and SiN (270 nm) layers remain unchanged. This treatment ensures that both the magnon-phonon and the magnon-photon coupling strength remain to be the same as those in the original-sized cavity for two reasons. First, the magnon-phonon coupling strength is determined only by the thicknesses of the YIG and SiN layer (along *y*) rather than their in-plane dimensions (*x* and *z*) because both the magnetization and strain vary only along the *y* axis. Second, the magnon-photon coupling strength is predominantly determined by the dimension ratio of the YIG-to-photon-cavity along the *x* axis because the cavity magnetic field $\mathbf{H}^{EM}$ varies largely along the *x* axis only (see Fig. 5a).

Because our focus is on the Kittel mode magnon and we are not interested in studying the size and shape effect of the YIG in this application example, we use one single cuboid-shaped cell with a size of ($\Delta x$, $\Delta y$, $\Delta z$) = (55 nm, 10 nm, 50 nm) to represent the YIG (i.e., the macrospin approximation) and a chain of 27 cells of the same dimension aligning along the $y$ axis to discretize the juxtaposed SiN layer. Cells of the same dimension are also used to discretize the reduced-sized photon cavity, resulting in a total number of cells ($N_x$, $N_y$, $N_z$)=(30, 40, 11). A planar source current $J_y^c(t) = J_y^{c,0} t e^{-t^2/2\sigma_0^2}$, which is spatially uniform along the $yz$ plane, is injected at $x$=275 nm to excite the cavity photon, with $J_y^{c,0}$=$10^{12}$ A/m$^2$ and $\sigma_0$=7×10$^{-11}$ s. The bias magnetic field $H_y^{\text{bias}}$ is applied along the +$y$ direction, causing the magnetization to precess around the $y$ axis, as shown in Fig. 5a. To set up the magnon-photon resonance, we identify the $H_y^{\text{bias}}$ that makes the FMR frequency $\omega_m$ equal the $\omega_c$=9.1 GHz by numerically simulating the mode splitting spectra of the magnon polaritons as a function of the $H_y^{\text{bias}}$ in the absence of coupling to phonons. A magnon-photon coupling strength $g_{cm}$ of 2π×56 MHz is obtained from the mode splitting spectra (see Supplementary Fig. S6c in Supplementary Note 10). We have found that the $g_{cm}$ remains unchanged when using other $\varepsilon_r$ values to reduce the dimensions of the cavity along $x$ and $z$ to other values (while fixing the $y$-axis cavity dimension to 400 nm) and proportionally reducing the $x$-axis and $z$-axis dimension of the YIG/SiN bilayer (while fixing the $y$-axis dimension of the bilayer to 280 nm). The results are shown in Supplementary Fig. S6d in Supplementary Note 10.

Regarding the set-up of the phonon resonator, since the acoustic wave is spatially uniform in the $xz$ plane, we use one single cell to represent the YIG/SiN bilayer along their $x$ and $z$ axes. Considering that the phonons are excited by a uniform magnetization precessing around the $y$ axis and omitting the elastic damping ($\beta$=0), Eq. (4) can be written as,

$$\rho \frac{\partial^2 u_x}{\partial t^2} = c_{44} \frac{\partial^2 u_x}{\partial y^2} + B_2 \frac{\partial (m_x m_y)}{\partial y}, \qquad (9)$$

$$\rho \frac{\partial^2 u_y}{\partial t^2} = c_{11} \frac{\partial^2 u_y}{\partial y^2} + B_1 \frac{\partial (m_y^2)}{\partial y}, \qquad (10)$$

$$\rho \frac{\partial^2 u_z}{\partial t^2} = c_{44} \frac{\partial^2 u_z}{\partial y^2} + B_2 \frac{\partial (m_y m_z)}{\partial y}, \qquad (11)$$

where $B_1$ and $B_2$ are the magnetoelastic coupling coefficients of the YIG resonator. Equations (8a-c) indicate that the Kittel mode magnons will excite chiral TA phonons of the same frequency which have a wavevector along the $y$ axis (see Fig. 5d). Using procedures similarly to those described in ref. [39], we analytically derive the frequencies of the standing TA phonon frequencies in the YIG/SiN bilayer as a function of the YIG and SiN layer thickness ($d^{\text{YIG}}$ and $d^{\text{SiN}}$), i.e.,

$$\frac{c_{44}^{\text{YIG}}}{v^{\text{YIG}}} \left(1 + e^{2i\omega_n \frac{d^{\text{SiN}}}{v^{\text{SiN}}}}\right) \left(-1 + e^{2i\omega_n \frac{d^{\text{YIG}}}{v^{\text{YIG}}}}\right) + \frac{c_{44}^{\text{SiN}}}{v^{\text{SiN}}} \left(-1 + e^{2i\omega_n \frac{d^{\text{SiN}}}{v^{\text{SiN}}}}\right) \left(1 + e^{2i\omega_n \frac{d^{\text{YIG}}}{v^{\text{YIG}}}}\right) = 0 \quad (12)$$

The first nonzero nontrivial solution of Eq. (12) yields the angular frequency of the fundamental ($n$ =1) acoustic phonon mode ($\omega_{n=1}$), and so forth for the higher-order modes. Here $v^{\text{YIG}} = \sqrt{c_{44}^{\text{YIG}}/\rho^{\text{YIG}}}$ is the velocity of the TA phonon in YIG and likewise for the $v^{\text{SiN}}$. When $d^{\text{YIG}}$= 10 nm and $d^{\text{SiN}}$= 270 nm, the resonant acoustic frequency $\omega_{n=1}/2\pi$=9.13 GHz, which is very close to the frequency (9.14 GHz) of the $\hat{d}_+$ mode magnon polariton. The elastic parameters of the YIG and SiN are provided in Supplementary Note 8.

**Data Availability**
The data that support the plots presented in this paper are available from the corresponding authors upon reasonable request.

**Code Availability**
Open-source codes for the present dynamical phase-field model can be accessed via https://github.com/jhu238/GO-Ferro.

**Acknowledgement**
J.-M.H. acknowledges the support from the National Science Foundation (NSF) under the grant number CBET-2006028 and DMR-2237884. The benchmarking test of the 3D elastodynamic solver of the dynamical phase-field model as well as the design of the cavity electro-magno-mechanical system with triple phonon-magnon-photon resonance are supported by the US Department of Energy, Office of Science, Basic Energy Sciences, under Award Number DE-SC0020145 as part of the Computational Materials Sciences Program. The dynamical phase-field simulations were performed using Bridges at the Pittsburgh Supercomputing Center through allocation TG-DMR180076 from the Advanced Cyberinfrastructure Coordination Ecosystem: Services & Support (ACCESS) program, which is supported by NSF grants #2138259, #2138286, #2138307, #2137603, and #2138296. X.Z. acknowledges support from the NSF (2337713) and ONR Young Investigator Program (N00014-23-1-2144). C.Z. and L.J. acknowledge support from the AFRL (FA8649-21-P-0781), NSF (ERC-1941583, OMA-2137642) and Packard Foundation (2020-71479).

**Author contributions**
J.-M.H. and X.Z. initiated the project and designed the structure of the paper. J.-M.H., L.J. and X.Z. supervised the research. S.Z. developed the computer codes for the dynamical phase-field model. S.Z. and Y.Z. performed the dynamical phase-field simulations. Y.Z. performed the benchmarking test of the 3D elastodynamic solver. C.Z. performed the Floquet Hamiltonian based theoretical calculations. J.-M.H. and S.Z. wrote the paper using substantial feedback from X.Z., C.Z., and L.J.. All authors contributed to data analysis.

**Competing Interests**
The authors declare no competing interests.

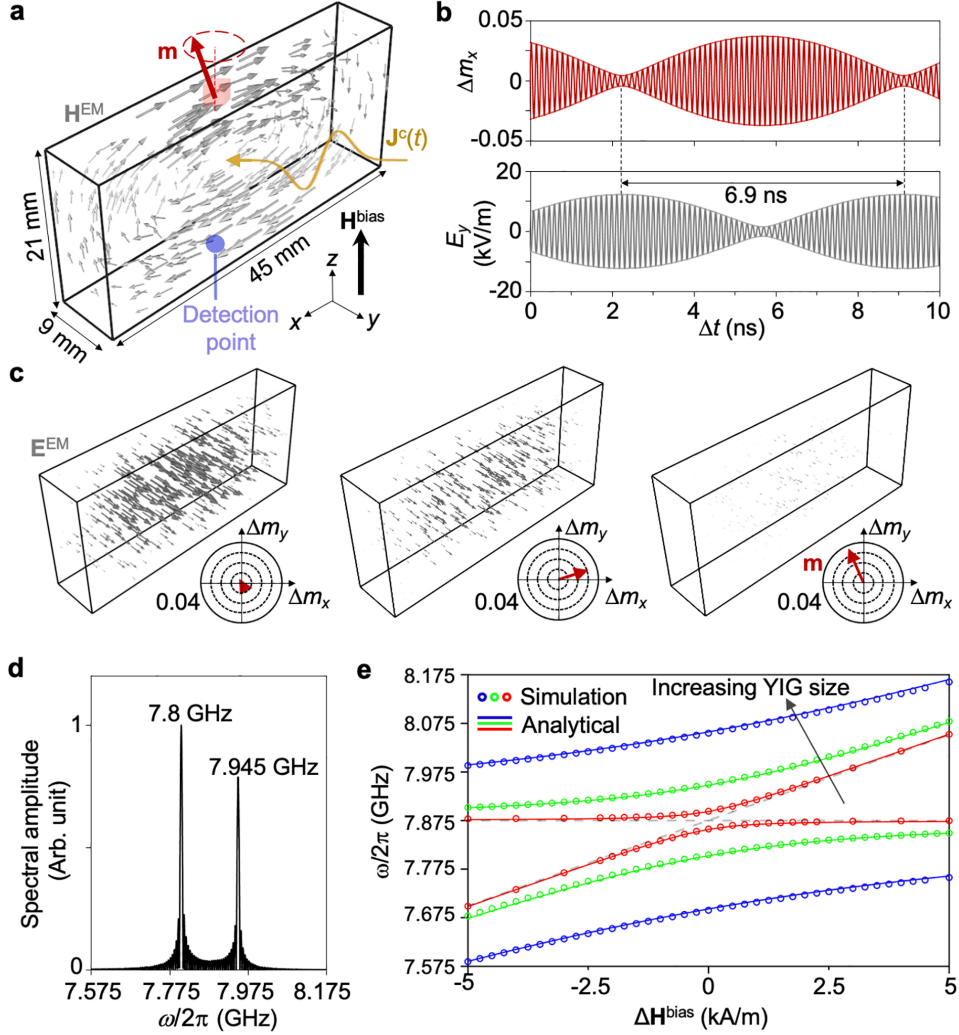

**Figure 1**. **The excitation of magnon polaritons. a**, Hybrid magnon-photonic system that contains a YIG cube (not to scale) inside a 3D microwave cavity. The bias magnetic field $\mathbf{H}^{bias}$ is applied along the $+z$ direction. A Gaussian-shaped charge current pulse $\mathbf{J}^c(t)$ is injected into the cavity to excite the cavity mode of the standing EM wave. The vectors indicate the direction of the local microwave magnetic field $\mathbf{H}^{EM}$ ($TE_{101}$ mode), where the vector length is proportional to magnitude of the $\mathbf{H}^{EM}$. **b**, Dynamics of the on-resonance Kittel mode magnon, represented by $\Delta m_x = m_x(t)-m_x(t=0)$) and cavity photon, represented by the microwave electric field component $E_y^{EM}$ at the detection point which is at 1.5 mm above the bottom surface center of the cavity. Trend lines showing the evolution of the amplitudes of the two modes are added. The values of the simulated $E_y^{EM}$ are multiplied by $\sqrt{\varepsilon_r}$ to show the electric fields in the cavity of the original size. **c**, Spatial distribution of local microwave electric field $\mathbf{E}^{EM}$ and the polar plot of the magnitude of the precessing magnetization component $\Delta m_x$ and $\Delta m_y$ at (from left to right) $\Delta t$ = 2.11 ns, 4.4 ns, and 5.71 ns, respectively. The circles indicate $|\Delta \mathbf{m}|$=0.01 (innermost), 0.02, 0.03, and 0.04 (outermost), respectively. **d**, Frequency spectrum of the $\Delta m_x(t)$ in (**c**). **e**, Simulated mode splitting spectra of the magnon polaritons as a function of $\Delta \mathbf{H}^{bias}$ under different YIG sizes of 0.4×0.4×0.4 $mm^3$ (red circles), 1×1×1 $mm^3$ (green circles) and 2×2×2 $mm^3$ (blue circles), and their analytical fitting curves (lines). $\Delta \mathbf{H}^{bias} = \mathbf{H}^{bias} - \mathbf{H}_0^{bias}$, where $\mathbf{H}_0^{bias}$ = 0.2315 MA/m (~0.291 T) is the bias magnetic field that ensures magnon-photon on-resonance ($\omega_m = \omega_c$).

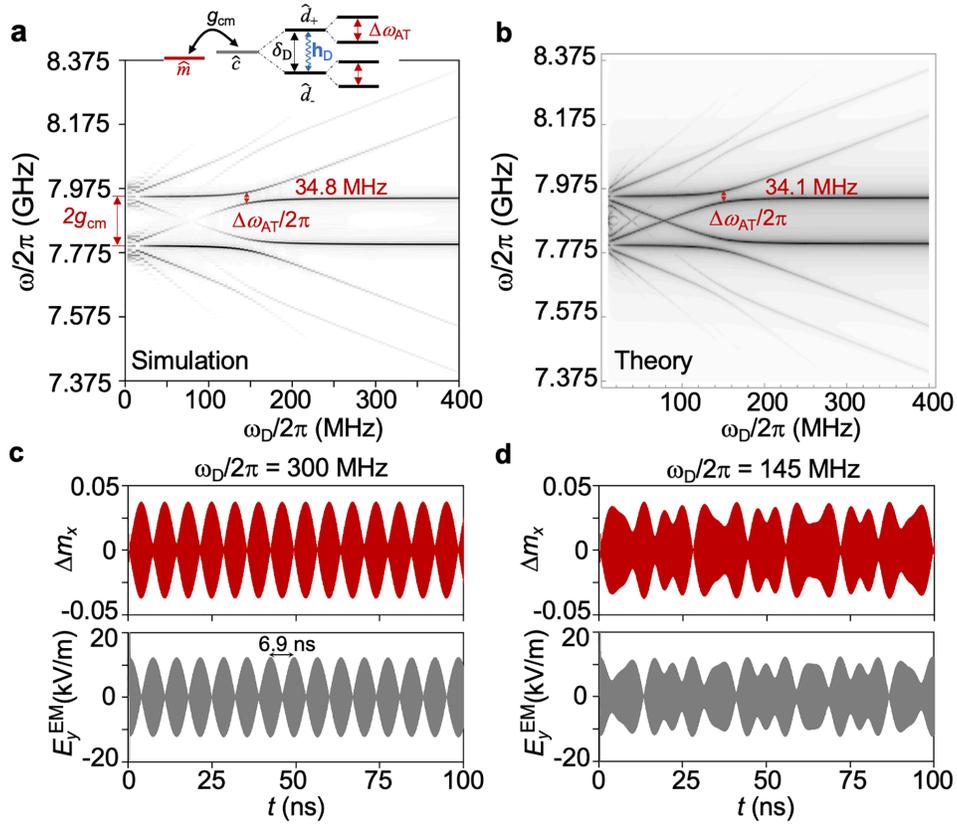

**Figure 2. Floquet-induced magnonic Aulter-Townes splitting. a**, Frequency spectrum of the magnon polariton as a function of Floquet driving frequency $f_D$, obtained from **a**, dynamical phase-field simulations and **b**, Hamiltonian-based theoretical calculations of the absorption spectrum where square root for each data point is take to make the higher order sideband visible. The inset shows the energy level diagram. Evolutions of the magnon (represented by $\Delta m_x$, upper panel) and the photon (represented by $E_y$, lower panel) at the detection point under **c**, $\omega_D/2\pi$=300 MHz, and **d**, $\omega_D/2\pi$=145 MHz. The magnitude of the Floquet driving field $|\mathbf{h}_D|$=2000 A/m in **a-d**.

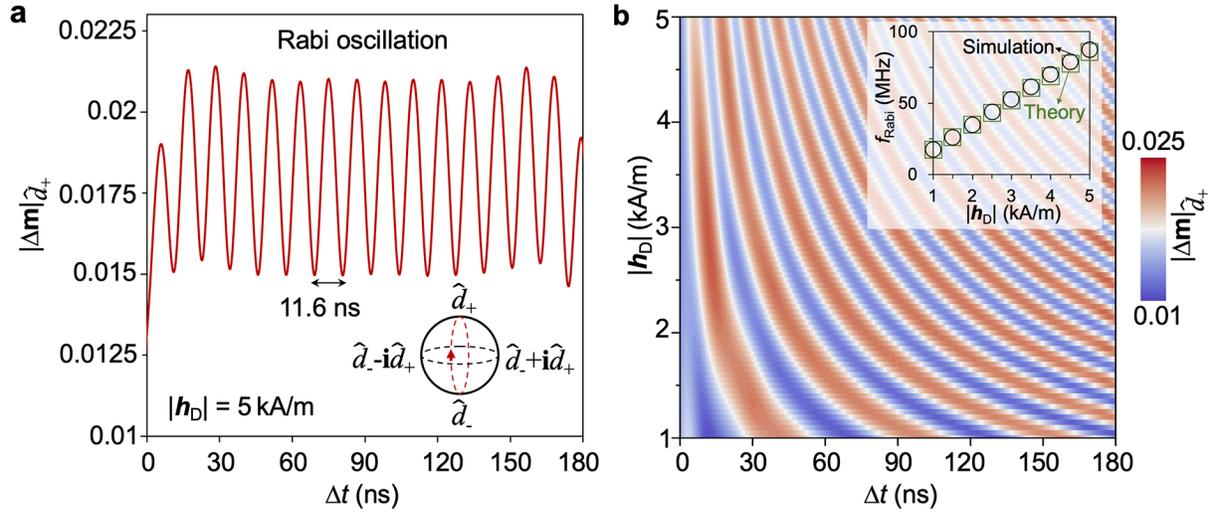

**Figure 3. Magnonic Rabi oscillation. a**, Dynamics of the magnetization amplitude of the $\hat{d}_+$ mode magnon polariton under a continuous Floquet driving $\mathbf{h}_D(t)$ with $|\mathbf{h}_D|$=5000 A/m and $\omega_D/2\pi$=145 MHz. $\Delta t$=0 is the moment when the application of $\mathbf{h}_D(t)$ begins after the current pulse $\mathbf{J}^c(t)$ injection is complete. The $\hat{d}_+$ and $\hat{d}_-$ modes swap at the frequency of $\Delta\omega_{AT}/2\pi$ along the real axis of the Bloch sphere (inset). **b**, Dynamics of the magnetization amplitude of the $\hat{d}_+$ mode magnon polariton under different $|\mathbf{h}_D|$ but the same frequency of $\omega_D/2\pi$=145 MHz.

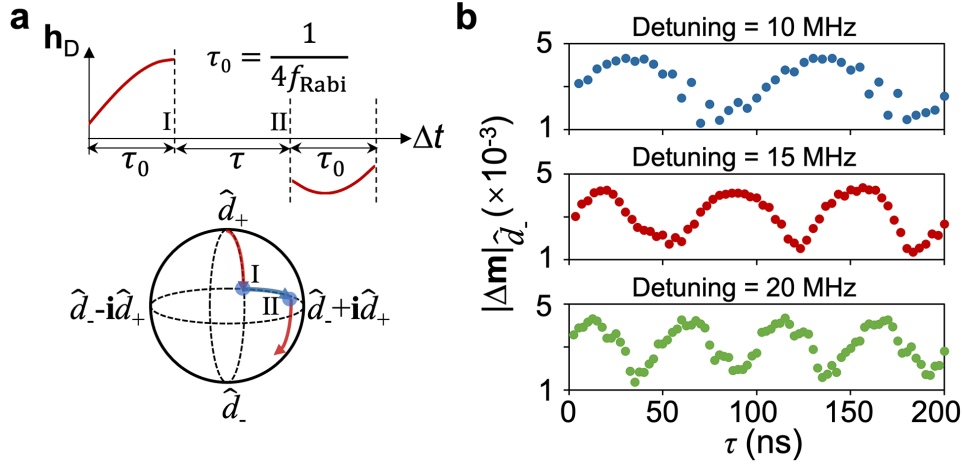

**Figure 4. Magnonic Ramsey interference. a.** (Top) the temporal waveform $\mathbf{h}_D(t)=|\mathbf{h}_D|\sin(\omega\Delta t)$ when $0 \leq \Delta t \leq \tau_0$ or $\tau_0+\tau \leq \Delta t \leq 2\tau+\tau_0$, and $\mathbf{h}_D(t)=0$ otherwise; (Bottom) Schematic of operation sequences for Ramsey interference on a Bloch sphere. **b**, The magnetization amplitude of the $\hat{d}_-$ mode of the magnon polariton obtained after the completion of the second $\pi/2$ pulse. Each data point in **b** was obtained from an independent simulation, where the free evolution duration $\tau$ and detuning amplitude $\Delta\omega$ are different in each simulation.

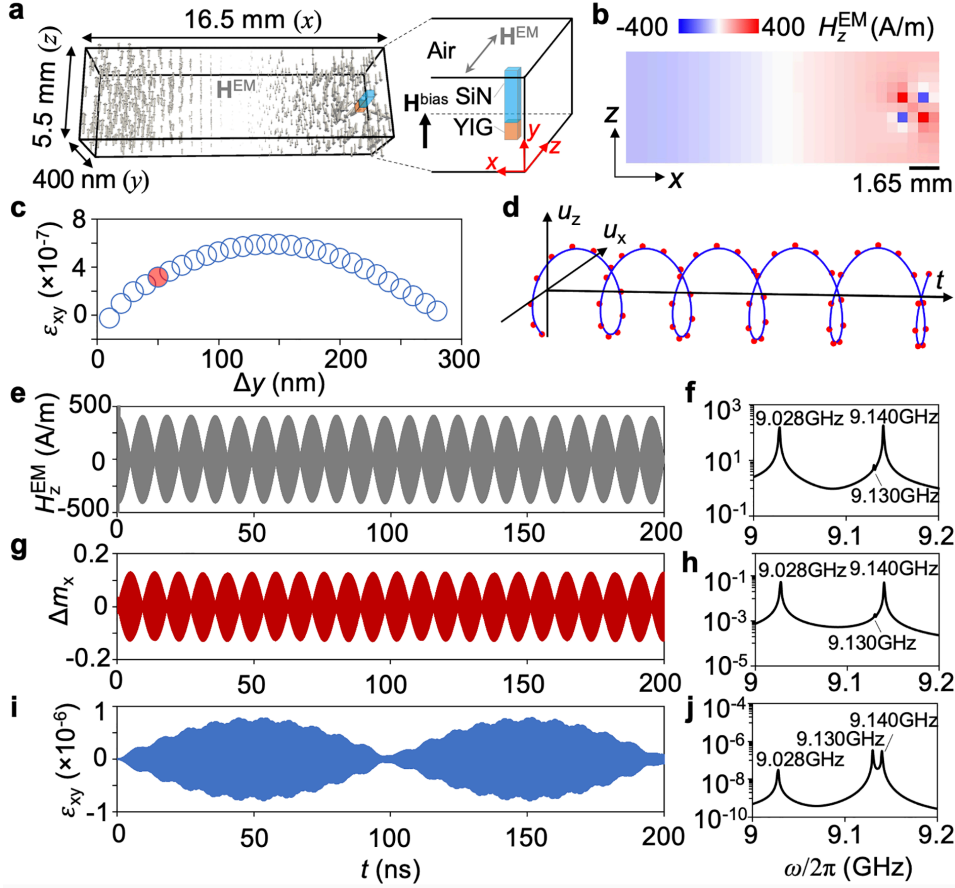

**Figure 5. Triple phonon-magnon-photon resonace. a**, A cavity electromagnonic system that contains a YIG/SiN bilayer membrane inside a 3D photon cavity. The arrow indicates the direction (mostly along $z$) and the vector length indicates the magnitude of the local cavity magnetic field $\mathbf{H}^{EM}$. The schematic on the right illustrates the YIG/SiN bilayer membrane (not to scale), which occupies the space where $x \in \{14.85$ mm, 15.4 mm$\}$, $y \in \{40$ mm, 280 nm$\}$, and $z \in \{2.5$ mm, 3 mm$\}$. The bias magnetic field $\mathbf{H}^{bias}$ is applied along $+y$. The lower left corner of the cavity is defined as coordinate of the origin, i.e., $(x, y, z) = (0,0,0)$. **b**, Spatial distribution of the $H_z^{EM}$ at $t=50$ ns in the $xz$ plane of the 3D system at $y=50$ nm. **c**, Profile of the strain component $\varepsilon_{xy}$ along the thickness direction ($y$) of the YIG/SiN bilayer membrane at $t=50$ ns. **d**, Evolution of the mechanical displacement $u_x$ and $u_z$ from $t=50$-$51$ ns at 50 nm above the bottom of the YIG/SiN bilayer membrane (this location is indicated by the filled circles in **c**). Under the triple phonon-magnon-photon resonance condition, evolution of, **e**, the TE$_{100}$ mode cavity photon, represented by the $H_z^{EM}$ at the point $(x, y, z) = (1.65$ mm, 50 nm, 3 mm$)$, **g**, the Kittel mode magnon, represented by $\Delta m_x = m_x(t) - m_x(t=0)$, and, **i**, the standing chiral TA phonon mode at the detection point, represented by the $\varepsilon_{xy}$ at the point indicated by the filled circle in **c**. $t=0$ is the moment the planar current pulse is injected to the cavity. **f,h,j**, Frequency spectra of the cavity photon, the Kittel mode magnon, and the chiral TA phonon, respectively.

# Dynamical phase-field model of cavity electromagnonic systems

Shihao Zhuang,[1a)] Yujie Zhu[1a)], Changchun Zhong,[2] Liang Jiang,[2] Xufeng Zhang[3,4†], Jia-Mian Hu[1*]

[1]Department of Materials Science and Engineering, University of Wisconsin-Madison, Madison, WI, 53706, USA

[2]Pritzker School of Molecular Engineering, The University of Chicago, Chicago, IL 60637, USA

[3]Department of Electrical and Computer Engineering, Northeastern University, Boston, Massachusetts 02115, USA

[4]Department of Physics, Northeastern University, Boston, Massachusetts 02115, USA

E-mails:
[†]xu.zhang@northeastern.edu
[*]jhu238@wisc.edu

[a)]S.Z. and Y.Z. contribute equally to this work.

**Supplementary Note 1**. List of symbols for various modes and related quantities

**Supplementary Note 2**. $|\mathbf{h}_D|$-dependent magnonic Autler-Townes splitting.

**Supplementary Note 3**. Understanding $|\mathbf{h}_D|$-dependent magnon-photon coupling strength $\tilde{a}$

**Supplementary Note 4**. Dynamics of $|\Delta\mathbf{m}|_+$ and $|\Delta\mathbf{m}|_-$ in magnonic Rabi oscillation

**Supplementary Note 5**. Understanding magnonic Rabi oscillation

**Supplementary Note 6**. Understanding magnonic Ramesy interference

**Supplementary Note 7**. Coupled mode dynamics under the triple phonon-magnon-photon resonance with elastic damping

**Supplementary Note 8**. Benchmarking test of the 3D Elastodynamic solver

**Supplementary Note 9.** EM fields under the same excitation current but different $\varepsilon_r$

**Supplementary Note 10**. Supporting data for the system under the triple phonon-magnon-photon resonance condition

**Supplementary Note 1**. List of symbols for various modes and related quantities

| Symbol | Explanation |
| --- | --- |
| $\hat{c}$ and $\hat{m}$ | Cavity photon mode and Kittel magnon mode, respectively |
| $\omega_c$ and $\omega_m$ | Angular frequency of the cavity photon mode and Kittel magnon mode, respectively. |
| $g_{cm}$ | Coupling strength between the cavity photon mode and Kittel magnon mode |
| $\omega_0$ | Resonant angular frequency, $\omega_0 = \omega_c = \omega_m$ |
| $\hat{d}_+$ and $\hat{d}_-$ | High and low-frequency mode of the magnon polaritons |
| $\omega_+$ and $\omega_-$ | Angular frequency of the $\hat{d}_+$ and $\hat{d}_-$ mode, respectively |
| $\omega_D$ | Angular frequency of the Floquet drive |
| $\Omega$ and $|\mathbf{h}_D|$ | $\Omega = \gamma|\mathbf{h}_D|$, where $|\mathbf{h}_D|$ is the amplitude of the Floquet drive |
| $\Delta\omega_{AT}$ | Angular frequency of magnonic Autler-Townes (AT) splitting |
| $f_{Rabi}$ | Rabi flopping frequency, which is found to be equal to $\Delta\omega_{AT}/2\pi$ |
| $\delta_D, \Delta\omega$ | $\delta_D = \omega_+ - \omega_- = 2g_{cm}$ is the cavity-magnon mode splitting frequency. $\Delta\omega = \omega_D - \delta_D$ is the frequency detuning |
| $\tau_0, \tau$ | $\tau_0$ is the duration of the $\pi/2$ pulse and $\tau$ is the free evolution duration in the Ramsey interference study |
| $\alpha$ | Initial amplitude of the $\hat{d}_+$ mode in the Ramsey interference study |

**Supplementary Note 2**. $|\mathbf{h}_D|$-dependent magnon-photon coupling strength and magnonic Autler-Townes splitting.

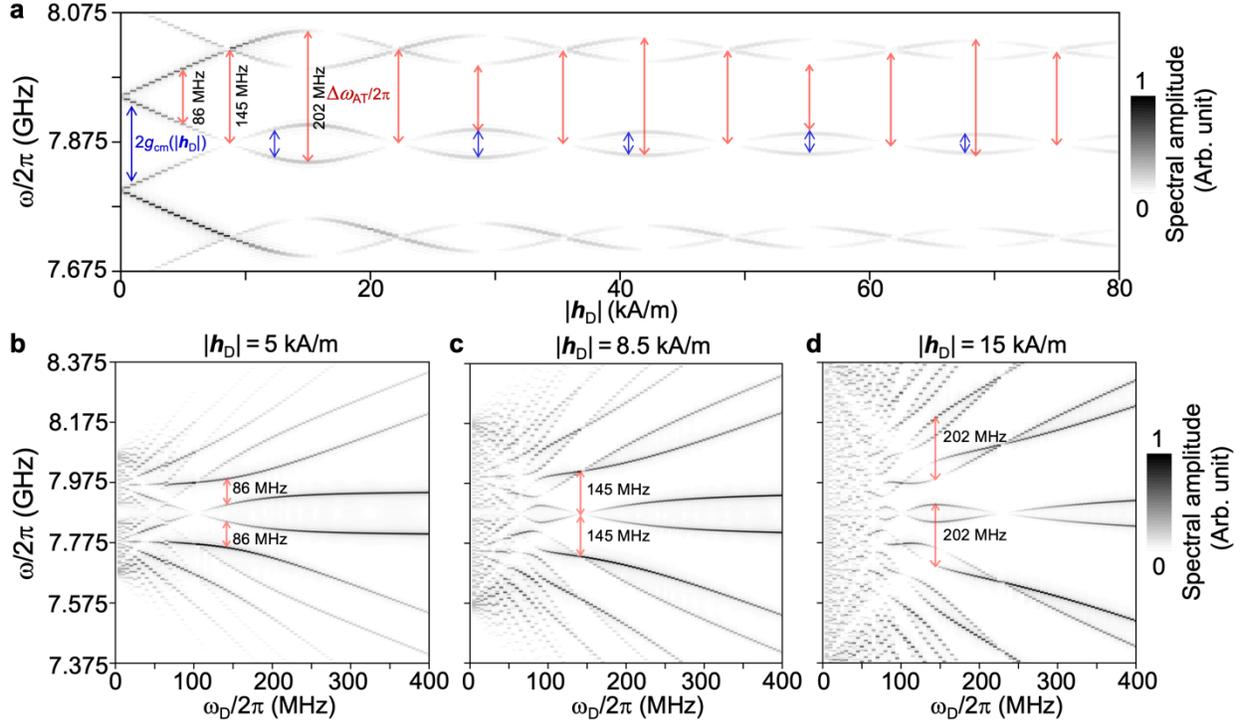

**Supplementary Figure S1**. **a**, Floquet-drive induced frequency splitting in the frequency spectrum of the magnon polariton as a function of the driving amplitude $|\mathbf{h}_D|$, under the same driving frequency $\omega_D/2\pi = \delta_D/2\pi$=145 MHz. Specifically, the figure presents the frequency spectra of the $\Delta m_x(t)$ simulated under different $|\mathbf{h}_D|$ but the same $\omega_D$. Frequency spectra of the magnon polariton as a function of Floquet driving frequency $f_D$ under the driving amplitude **b**, $|\mathbf{h}_D|$=5 kA/m, **c**, $|\mathbf{h}_D|$=8.5 kA/m and **d**, $|\mathbf{h}_D|$=15 kA/m. Likewise, these are obtained by performing Fourier transform of the $\Delta m_x(t)$ data. The red arrows in **a** indicate the magnonic Autler-Townes splitting with a frequency of $\Delta\omega_{AT}/2\pi$=86 MHz, 145 MHz, and 202 MHz at $|\mathbf{h}_D|$=5 kA/m, 8.5 kA/m, and 15 kA/m, respectively. When $|\mathbf{h}_D|$ exceeds 8.5 kA/m, $\Delta\omega_{AT} > 2g_{cm}$, where the system is in the Floquet ultrastrong coupling regime. The blue arrows indicate that the $2g_{cm}$ varies periodically (down to 0) with $|\mathbf{h}_D|$ in an oscillatory manner. The red arrows indicate that the variation of $\Delta\omega_{AT}$ with $|\mathbf{h}_D|$ is also oscillatory and periodic, but the values of $\Delta\omega_{AT}$ remains nonzero at large $|\mathbf{h}_D|$. The variation trends of both the $g_{cm}$ and the $\Delta\omega_{AT}$ can be understood by deriving their analytical formulae as a function of $\mathbf{h}_D$, as shown in Supplementary Note 3.

# Supplemental Material 3: Understanding $|h_D|$-dependent magnon-photon coupling strength

As shown in Supplemental Material 2 (Fig. S1), the coupling strength between the magnon mode an the cavity photon mode varies in an oscillatory manner as the strength of the Floquet drive increases. To quantitatively understand this phenomenon, we start by writing,

$$\hat{H} = \omega_c \hat{c}^\dagger \hat{c} + \omega_m \hat{m}^\dagger \hat{m} + g_{cm}(\hat{c}^\dagger \hat{m} + \hat{c}\hat{m}^\dagger) + \Omega \hat{m}^\dagger \hat{m} \cos(\omega_D t). \tag{1}$$

This Hamiltonian can be expressed in the following rotating frame $U = e^{i\varepsilon(t)\hat{m}^\dagger \hat{m}}$ as

$$\hat{H} = \omega_c \hat{c}^\dagger \hat{c} + \omega_m \hat{m}^\dagger \hat{m} + g_{cm}\left[\hat{c}^\dagger \hat{m} e^{-i\varepsilon(t)} + \hat{c}\hat{m}^\dagger e^{i\varepsilon(t)}\right]. \tag{2}$$

The phase factor $\varphi(t) = e^{i\varepsilon(t)}$ can be expressed in the following series

$$\varphi(t) = \sum_{n=-\infty}^{\infty} J_n\left(\frac{\Omega}{\omega_D}\right) e^{-i\cdot n(\omega_D t)}. \tag{3}$$

Thus we have

$$\hat{H} = \omega_c \hat{c}^\dagger \hat{c} + \omega_m \hat{m}^\dagger \hat{m} + g_{cm} \sum_{n=-\infty}^{\infty} J_n\left(\frac{\Omega}{\omega_D}\right)\left[\hat{c}^\dagger \hat{m} e^{i\cdot n(\omega_D t)} + \hat{c}\hat{m}^\dagger e^{-i\cdot n(\omega_D t)}\right]. \tag{4}$$

From the above Hamiltonian, we see that the coupling strength is modified by the Bessel function. Specifically, if we look at the term with $n = 0$, we have

$$\hat{H} = \omega_c \hat{c}^\dagger \hat{c} + \omega_m \hat{m}^\dagger \hat{m} + g_{cm} J_0\left(\frac{\Omega}{\omega_D}\right)(\hat{c}^\dagger \hat{m} + \hat{c}\hat{m}^\dagger) + g_{cm} \sum_{n\neq 0} J_n\left(\frac{\Omega}{\omega_D}\right)\left[\hat{c}^\dagger \hat{m} e^{-i\cdot n(\omega_D t)} + \hat{c}\hat{m}^\dagger e^{i\cdot n(\omega_D t)}\right], \tag{5}$$

where the $|h_D|$-dependent coupling strength between the magnon and the cavity mode is obtained as,

$$g_{cm} = g_{cm}\left|J_0\left(\frac{\Omega}{\omega_D}\right)\right|, \tag{6}$$

where the $g_{cm}$ on the right-hand side is the coupling strength at $|h_D| = 0$, which is $2\pi \times 72.5 MHz$.

Similarly, we have in certain rotating frame

$$\hat{H} = (\omega_c + \omega_D)\hat{c}^\dagger \hat{c} + (\omega_m + \omega_D)\hat{m}^\dagger \hat{m} + g_{cm} J_0\left(\frac{\Omega}{\omega_D}\right)(\hat{c}^\dagger \hat{m} + \hat{c}\hat{m}^\dagger) + g_{cm} \sum_{n\neq 0} J_n\left(\frac{\Omega}{\omega_D}\right)\left[\hat{c}^\dagger \hat{m} e^{-i\cdot n(\omega_D t)} + \hat{c}\hat{m}^\dagger e^{i\cdot n(\omega_D t)}\right], \tag{7}$$

which explains the repeated curve in terms of driving strength. The corresponding gaps between different bands can be obtained accordingly. Moreover, the $|h_D|$-dependent magnonic Autler-Townes splitting $\Delta\omega_{AT}$ can be analytically calculated as,

$$\Delta\omega_{AT} = 2g_{cm}\left[1 - J_0\left(\frac{\Omega}{\omega_D}\right)\right], \tag{8}$$

where the $g_{cm}$ on the right-hand side is likewise the coupling strength at $|h_D| = 0$.

**Supplementary Note 4.** Dynamics of $|\Delta\mathbf{m}|_{\hat{a}_+}$ and $|\Delta\mathbf{m}|_{\hat{a}_-}$ in magnonic Rabi oscillation

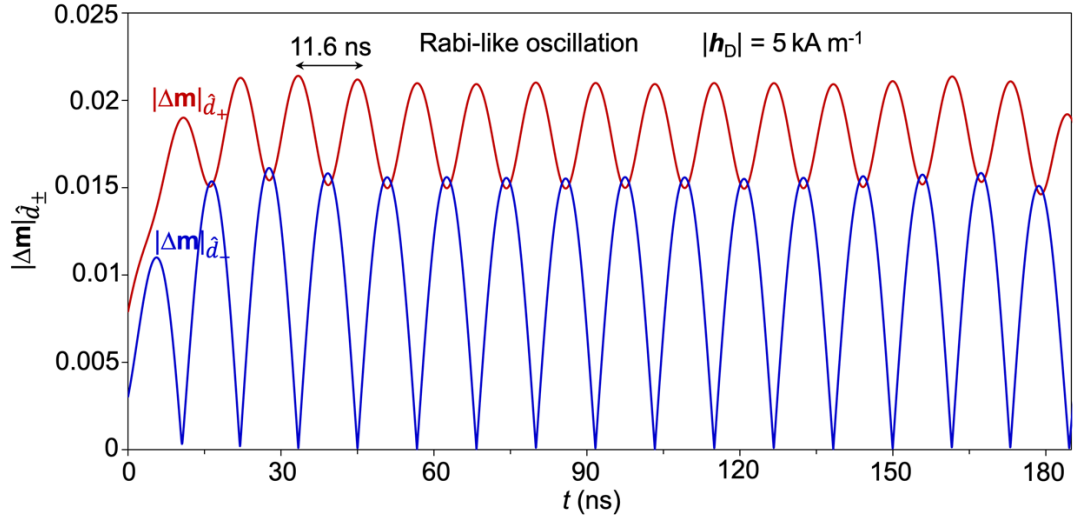

**Supplementary Figure S2**. Dynamics of the numerically extracted magnetization amplitude of the $\hat{a}_+$ and $\hat{a}_-$ magnon polariton mode in the initiation stage ($t=0\sim1.89$ ns) where a 15-cycle $\mathbf{J}^c(t)=J_0\sin(\omega t)$ at $\omega=2\pi\times7.945$ GHz along the $y$ axis (*i.e.*, only the $J_y^c$ component is nonzero) to populate (excite) the $\hat{a}_+$ mode, and then in the Floquet drive stage ($t>1.89$ ns) where the current is turned off but a continuous Floquet drive $\mathbf{h}_D(t)$ at the cavity-magnon mode splitting frequency ($\omega_D/2\pi=\delta_D/2\pi=145$ MHz) is applied to the system.

# Supplemental Material 5: Understanding magnonic Rabi oscillation

Initially, we have

$$\hat{H} = \omega_c \hat{c}^\dagger \hat{c} + \omega_m \hat{m}^\dagger \hat{m} + g_{cm}(\hat{c}^\dagger \hat{m} + \hat{c}\hat{m}^\dagger) + \Omega \hat{m}^\dagger \hat{m} \cos(\omega_D t). \quad (9)$$

Every terms should be self-evident. Take $\omega_c = \omega_m = \omega_0$. Now we define two hybridized modes given the strong coupling condition,

$$\hat{d}_+ = (\hat{c} + \hat{m})/\sqrt{2}, \hat{d}_- = (\hat{c} - \hat{m})/\sqrt{2}.$$

Define the rotating frame transformation

$$U = e^{-i\hat{d}_+^\dagger \hat{d}_+ \varepsilon(t)/2 - i\hat{d}_-^\dagger \hat{d}_- \varepsilon(t)/2},$$

where

$$\varepsilon(t) = \frac{\Omega}{\omega_D}[\sin(\omega_D t)].$$

Then we obtain the following Hamiltonian in that rotating frame

$$\hat{H} = \omega_+ \hat{d}_+^\dagger \hat{d}_+ + \omega_- \hat{d}_-^\dagger \hat{d}_- + \frac{\Omega}{2}\cos(\omega_D t)(\hat{d}_+^\dagger \hat{d}_- + \hat{d}_+ \hat{d}_-^\dagger), \quad (10)$$

where the mode frequency $\omega_\pm = \omega_0 \pm g_{cm}$. Each mode follow the the Heisenberg equation

$$\frac{d}{dt}\begin{pmatrix}\hat{d}_+ \\ \hat{d}_-\end{pmatrix} = -i\begin{pmatrix}\omega_+ & \Omega/2\cos(\omega_D t) \\ \Omega/2\cos(\omega_D t) & \omega_-\end{pmatrix}\begin{pmatrix}\hat{d}_+ \\ \hat{d}_-\end{pmatrix} \quad (11)$$

The above equations cannot be solved exactly. Instead, we solve it with a rotating wave approximation. Let's define

$$\hat{d}_+ \equiv \hat{d}_+ e^{-i\omega_+ t}, \hat{d}_1 \equiv \hat{d}_1 e^{-i\omega_- t}. \quad (12)$$

We get

$$\frac{d}{dt}\begin{pmatrix}\hat{d}_+ \\ \hat{d}_-\end{pmatrix} = -i\begin{pmatrix}0 & \Omega/2\cos(\omega_D t)e^{i\delta_D t} \\ \Omega/2\cos(\omega_D t)e^{-i\delta_D t} & 0\end{pmatrix}\begin{pmatrix}\hat{d}_+ \\ \hat{d}_-\end{pmatrix}$$

$$\simeq -i\begin{pmatrix}0 & \frac{\Omega}{4}e^{-i(\omega_D - \delta_D)t} \\ \frac{\Omega}{4}e^{i(\omega_D - \delta_D)t} & 0\end{pmatrix}\begin{pmatrix}\hat{d}_+ \\ \hat{d}_-\end{pmatrix} \quad (13)$$

where $\delta_D = \omega_+ - \omega_- = 2g_{cm}$. In the Rabi experiment, we have $\omega_D = \delta_D$, and initially we only populate the $\hat{d}_+$ mode, e.g., $\hat{d}_+(0) = 1$. The above equation group can be easily solved as

$$d_+(t) = \cos(\frac{\Omega}{4}t), d_-(t) = -i\sin(\frac{\Omega}{4}t). \quad (14)$$

Thus the mode square $|\hat{d}_+(t)|^2 = \cos^2(\frac{\Omega}{4}t)$, where $\Omega/2$ give the Rabi flopping frequency. Since we have $\Omega = \gamma|h_D|$, given $h_D = 5kA/m = 1.26 \times 10^{-3}T$ and $\gamma = 28GHz/T$, we have the Rabi flopping frequency

$$\Omega/2 = 176.4/2 = 88.2 MHz \quad (15)$$

which matches the experimental result.

## Supplemental Material 6: Understanding magnonic Ramsey interference

In the above equation, if the the drive is not equal to the hybridized mode difference $\omega_D \neq \delta_D$, we have the solution

$$d_+(t) = e^{-\frac{1}{2}i(\omega_D - \delta_D)t} \left( \cos(\frac{t}{4}\sqrt{\Omega^2 + 4(\omega_D - \delta_D)^2}) + \frac{2i(\omega_D - \delta_D)\sin(\frac{t}{4}\sqrt{\Omega^2 + 4(\omega_D - \delta_D)^2})}{\sqrt{\Omega^2 + (\omega_D - \delta_D)^2}} \right), \quad (16)$$

which means

$$|d_+(t)|^2 = \cos^2\left(\frac{t}{4}\sqrt{\Omega^2 + 4(\omega_D - \delta_D)^2}\right) + \frac{4(\omega_D - \delta_D)^2}{\Omega^2 + (\Omega_D - \delta_D)^2} \sin^2\left(\frac{t}{4}\sqrt{\Omega^2 + 4(\omega_D - \delta_D)^2}\right). \quad (17)$$

It gives a Rabi frequency $\sqrt{\Omega^2 + 4(\omega_D - \delta_D)^2}/2$. It is worth noting that the $\pi/2$ pulse must be defined with pump strength dependence in later Ramsey test.

The Eq. 13 can be rewritten as the following time independent form

$$\frac{d}{dt}\begin{pmatrix}\hat{d}_+ \\ \hat{d}_-\end{pmatrix} \simeq -i \begin{pmatrix} \frac{\omega_D - \delta_D}{2} & \frac{\Omega}{4} \\ \frac{\Omega}{4} & -\frac{\omega_D - \delta_D}{2} \end{pmatrix}\begin{pmatrix}\hat{d}_+ \\ \hat{d}_-\end{pmatrix}. \quad (18)$$

Now it becomes quite clear that the modes $\hat{d}_+, \hat{d}_-$ have new effective modes frequencies

$$\omega_{e+} = \frac{\omega_D - \delta_D}{2}, \omega_{e-} = -\frac{\omega_D - \delta_D}{2} \quad (19)$$

with the effective mode frequency separation $\Delta \omega = \omega_D - \delta_D$.

So the Ramsey process becomes self-evident. The first $\pi/2$ pulse prepares the state at the equator, then accumulate the phase difference $\Delta f \times \tau$ with the time delay $\tau$. The second $\pi/2$ pulse and the subsequent measurement reveals the oscillation

$$|d_+(\tau)|^2 = \frac{|\alpha|^2}{2}\cos^2\left(\frac{\Delta \omega \times \tau}{2}\right), \quad (20)$$

$$|d_-(\tau)|^2 = \frac{|\alpha|^2}{2}\sin^2\left(\frac{\Delta \omega \times \tau}{2}\right), \quad (21)$$

meanning the Ramsey frequency is exactly $\Delta \omega$.

**Supplementary Note 7. Coupled mode dynamics under the triple phonon-magnon-photon resonance with elastic damping.**

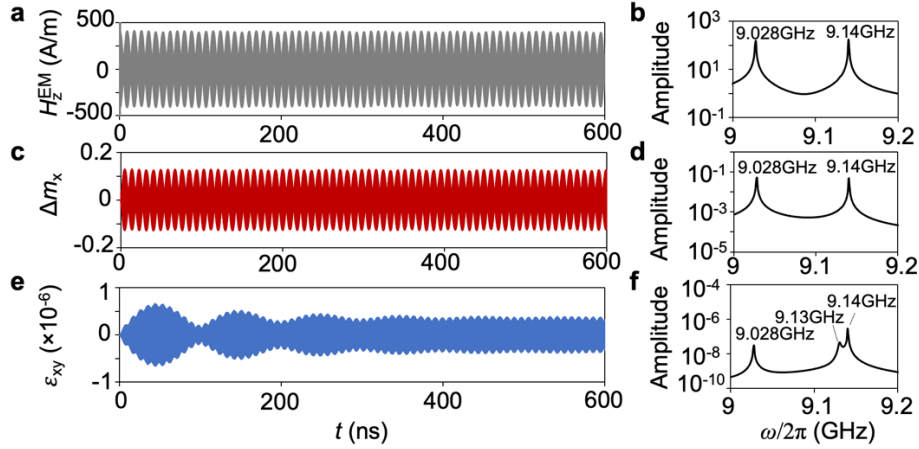

**Supplementary Figure S3**. Under the triple phonon-magnon-photon resonance condition, evolution of, **a**, the TE$_{100}$ mode cavity photon, represented by the $H_z^{EM}$ at the point $(x, y, z) = (1.65$ mm, 50 nm, 3 mm$)$, the same as that in Fig. 5e, **c**, the Kittel mode magnon, represented by $\Delta m_x = m_x(t) - m_x(t=0)$, and, **e**, the standing chiral TA phonon mode, represented by the $\varepsilon_{xy}$ at the point in the SiN (the location is the same as that in Fig. 5c). $t=0$ is the moment the planar current pulse is injected to the cavity. **b,d,f**, Frequency spectra of the time-domain data in **a,c,e**, respectively. An elastic stiffness damping coefficient $\beta = 4.48 \times 10^{-15}$ s is used for both the YIG and SiN, the same as the test in Supplementary Note 8.

## Supplementary Note 8. Benchmarking test of the 3D Elastodynamic solver

To demonstrate the high numerical accuracy of the in-house 3D elastodynamic solver, we consider a 3D discretized system of cuboid-shaped cells with a number of $N_x\Delta x \times N_y\Delta y \times N_z\Delta z$, where the cell size $(\Delta x, \Delta y, \Delta z) = (2\ \text{nm}, 2\ \text{nm}, 0.5\ \text{nm})$ and the number of cells $N_x=N_y=220$, and $N_z=45$. As shown in Fig. S4a, The YIG cube and the adjacent SiN membrane are surrounded by air. Specifically, the cells where $N_x \in \{61, 160\}$, $N_y \in \{61, 160\}$, and $N_z \in \{16, 40\}$ are designated as the YIG, the cells where $N_x \in \{11, 210\}$, $N_y \in \{11, 210\}$, and $N_z \in \{6, 15\}$ are designated as the SiN, and the remaining cells are designated as the air. The continuity boundary condition for $\mathbf{u}$ and $\boldsymbol{\sigma}$ are applied to all heterointerfaces. In this regard, by setting the $c_{ijkl}$ of the air to be zero, the stress-free surfaces of the SiN and YIG, at which the elastic waves would be reflected, are automatically considered. A Gaussian-shaped stress pulse $\sigma_{zz}(N_z=40, t) = \sigma_{\max} e^{-(t-5\tau)^2/2\tau^2}$ is applied at the top surface of the YIG cube at $t=0$ ps, where $\sigma_{\max}=3$ GPa and $\tau=1.5$ ps. The elastodynamic equation (Eq. (3) in the main text) is then solved using the classical Runge-Kutta method with $\Delta t = 2\times 10^{-15}$ s for time-marching and the central finite difference for calculating spatial derivatives. The outputs of the solver are the spatiotemporal evolution of the $\mathbf{u}$ and the resulting dynamical strain $\Delta\varepsilon_{ij}$.

In the COMSOL Multiphysics, the structural mechanics module was used to simulate the elastic wave propagation in a multiphase heterostructure. To reduce the computational cost, we consider a 3D structure of YIG and SiN that is one-fourth of the 3D structure in Fig. S4a, apply symmetry boundary condition to the inner lateral surfaces of the structure, and use cuboid-shaped cells with a size of $(\Delta x, \Delta y, \Delta z)=(2\ \text{nm}, 2\ \text{nm}, 0.5\ \text{nm})$ for the meshing, as shown in Fig. S4b. The YIG is confined to the domain where $x \in \{100\ \text{nm}, 200\ \text{nm}\}$, $y \in \{100\ \text{nm}, 200\ \text{nm}\}$, and $z \in \{5\ \text{nm}, 17.5\ \text{nm}\}$. The SiN is confined to the domain where $x \in \{0, 200\ \text{nm}\}$, $y \in \{0, 200\ \text{nm}\}$, and $z \in \{0, 5\ \text{nm}\}$. The backward Euler method is used for the time marching, with a fixed time step of $\Delta t = 5\times 10^{-14}$ s. At the top surface of the YIG ($z = 17.5$ nm), an Gaussian-shaped stress pulse $\sigma_{zz}(z=17.5\ \text{nm}, t) = \sigma_{\max} e^{-(t-5\tau)^2/2\tau^2}$ is applied, where $\sigma_{\max}=3$ GPa and $\tau=1.5$ ps, the same as the in-house solver. All the other surfaces are treated as free surfaces, at which the strain waves would be reflected.

The elastic stiffness coefficients $\mathbf{c}$, the mass density $\rho$, and the stiffness damping coefficient $\beta$ used in the simulations are listed as follows. For YIG (taken from ref. 39 of the main paper), $c_{11}$=269 GPa, $c_{12}$=107.7 GPa, $c_{44}$=76.4 GPa, $\rho$=5170 kg/m³; For SiN (taken from ref. 41 of the main paper), $c_{11}$=283.81 GPa, $c_{12}$=110.37 GPa, $c_{44}$=86.72 GPa, $\rho$=3170 kg/m³. For simplicity, we used an identical phenomenological stiffness damping coefficient $\beta$=4.48×10⁻¹⁵ s (following ref. 41) for both the YIG and the SiN. As shown in Fig. S4, the evolution of the strain $\varepsilon_{xx}$, $\varepsilon_{yy}$, and $\varepsilon_{zz}$ at the two selected points is almost the same in the in-house solver and COMSOL. We also postulate that the precision of the in-house solver is higher in this case due to (i) the use of a smaller time step $\Delta t$ and (ii) the Runge-Kutta method has a higher accuracy (albeit being more time-assuming) than the backward Euler method in terms of time marching.

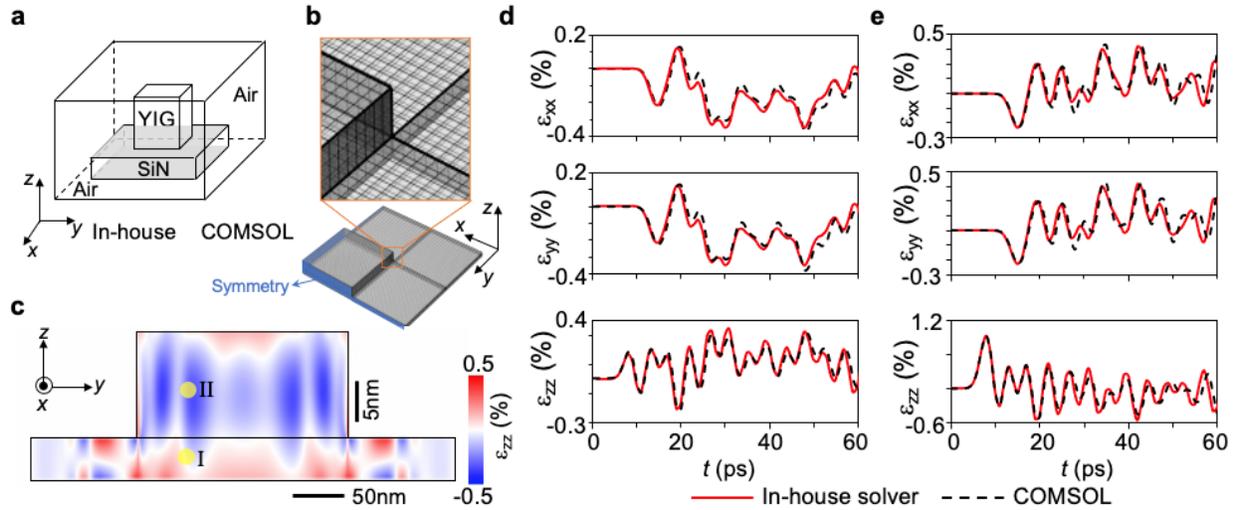

**Supplementary Figure S4**. Schematics of the 3D structure model in **a**, the in-house elastodynamic solver, and **b**, the COMSOL Multiphysics. In COMSOL, the origin (0 nm, 0 nm, 0 nm) is at the top right corner at the bottom surface of the SiN. **c**, Distribution of the strain $\varepsilon_{zz}$ in the 2D slice (i.e., the y-z plane) of the 3D model at $N_x$=110 at $t$=20 ps, obtained by the in-house solver. Evolution of the strain $\varepsilon_{xx}$, $\varepsilon_{yy}$, and $\varepsilon_{zz}$ at **d**, point I which corresponds to the $(N_x, N_y, N_z)$= (85, 85, 10) in the 3D structure for the in-house solver and $(x,y,z)$= (149 nm, 149 nm, and 2.25 nm) for the COMSOL model, and **e**, point II which corresponds to the $(N_x, N_y, N_z)$= (85, 85, 28) in the 3D structure for the in-house solver and $(x,y,z)$= (149 nm, 149 nm, and 11.25 nm) for the COMSOL model. $t$=0 ps is the moment that stress pulse is applied to the top surface of the YIG cube.

**Supplementary Note 9. EM fields under the same excitation current but different $\varepsilon_r$.**

In Fig. 1, the size of the cavity is 45×9×21 mm³ in the case of $\varepsilon_r$=1. We fix the number of cells for 3D cavity electromagnonic system with $N_x\Delta x \times N_y\Delta y \times N_z\Delta z$, with ($N_x$, $N_y$, $N_z$)= (90, 18, 42), leading to a cell size $\Delta x=\Delta y=\Delta z=l$=0.5 mm. YIG resonator occupies the domain at $N_x \in \{45, 46\}$, $N_y \in \{9, 10\}$, $N_z \in \{39, 40\}$. As the $\varepsilon_r$ increases, the EM wavelength is reduced by $\sqrt{\varepsilon_r}$ times. To describe the reduced EM wavelength, we reduce the cell size $l$ to $1/\sqrt{\varepsilon_r}$ of its value at $\varepsilon_r$=1. Specifically, as the $\varepsilon_r$ increases from 1 to $1\times10^{10}$, $2.25\times10^{10}$, and $4\times10^{10}$, the cell size $l$ is reduced from 0.5 mm to 5 nm, 3.33 nm, and 2.5 nm, respectively.

In our model set-up, the excitation current density $J_c$ (unit: C m⁻² s⁻¹) was applied by applying a time-varying point charge density $\rho_c = \rho_0 \left(\frac{t-5\sigma_0}{\sigma_0}\right) e^{-\left(\frac{t-5\sigma_0}{\sigma_0}\right)^2}$ with a unit of C m⁻³ s⁻¹ to the cell at ($N_x$, $N_y$, $N_z$) = (45, 3, 21), with $J_c=\rho_c l$ and $\sigma_0$ =70 ps. To ensure the magnitude of $J_c$ remains unchanged for different cases of $\varepsilon_r$, we increase the amplitude of $\rho_c$ by $\sqrt{\varepsilon_r}$ times for each case of $\varepsilon_r$. We have found that the $\Delta m_x(t)$ and $H_x(t)$ simulated under different values of $\varepsilon_r$ are identical, demonstrating that this scaling strategy in simulating the coupled dynamics of Kittel magnon mode and cavity photon mode is valid.

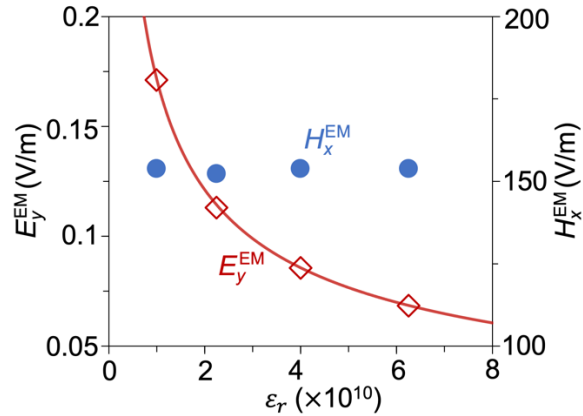

**Supplementary Figure S5.** The peak amplitudes of the electric-field and magnetic-field component of the EM wave ($E_y^{EM}$ and $H_x^{EM}$) at the cell ($N_x$, $N_y$, $N_z$) = (45, 9, 3) in the cavity within $t$=5 ns to 25 ns. $t$=0 is the moment the excitation current pulse $J_c$ is injected. The solid line is the fitting based on the $1/\sqrt{\varepsilon_r}$.

**Supplementary Note 10. Supporting data for the system under the triple phonon-magnon-photon resonance condition.**

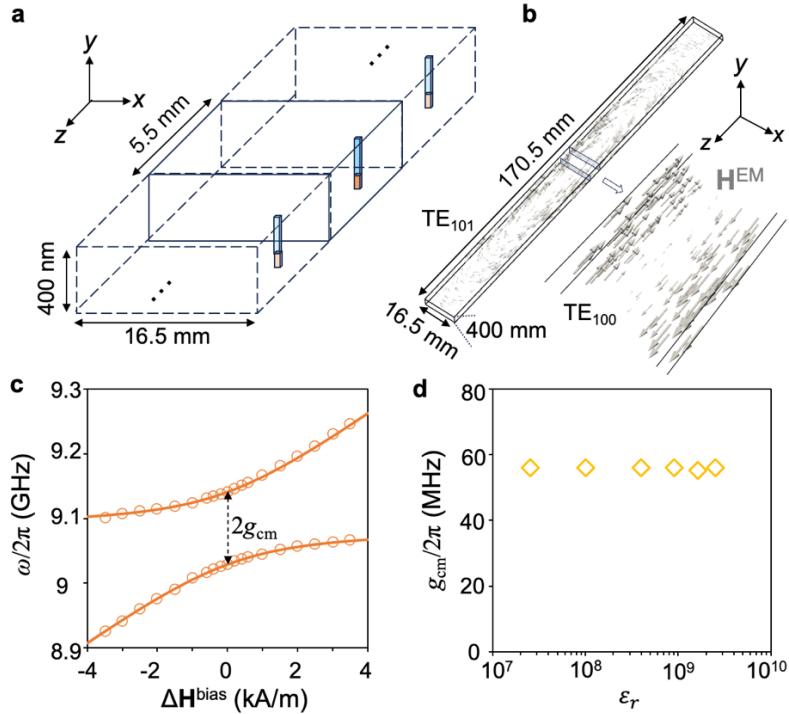

**Supplementary Figure S6. a**, Schematic of a full-scale cavity design that contains an array of YIG/SiN bilayer membranes (not to scale) placed in a 3D photon cavity. The number of repeating units (including the YIG/SiN bilayer) is larger than the three units shown in **a**. **b**, Distribution of the cavity magnetic field component in a larger-scale cavity, simulated using our dynamical phase-field model. The dimensions of the cavity along all three axes are indicated. PEC boundary condition is applied to all the surfaces of the photon cavity. Notably, although $TE_{101}$ mode is the lowest-order (preferrable) cavity photon mode, the cavity magnetic field in one repeating unit can be approximated as the $TE_{100}$ mode when the cavity length along the $z$ axis is long enough, as shown by the zoom-in figure in the bottom right panel. **c**, Simulated model splitting spectra (data indicated by the solid circles) of the magnon polaritons in the $TE_{100}$ cavity mode under $\varepsilon_r=10^8$ with a magnon-photon coupling strength $g_{cm}=2\pi\times56$ MHz, which is equal to half of the frequency gap at $\Delta H^{bias}=H^{bias}-H_0^{bias}=0$ (the gap is indicated by the double-headed arrow). The bias magnetic field is applied along the $+y$ direction of the photon cavity, with $H_0^{bias}=0.385$ MA/m. The solid line is the fitting curve using the analytical formula described in the main text, similarly to Fig. 1e. **d**, Simulated $g_{cm}$ (based on the $TE_{100}$ mode) as a function of the $\varepsilon_r$ in cavities that are down-scaled to $16.5/\sqrt{\varepsilon_r}$ mm ($x$) × 400 nm ($y$) × $5.5/\sqrt{\varepsilon_r}$ mm ($z$) along the $x$ and $z$ axis. The size of the YIG/SiN bilayer is proportionally reduced to $0.55/\sqrt{\varepsilon_r}$ mm ($x$) × 280 nm ($y$) × $0.5/\sqrt{\varepsilon_r}$ mm ($z$).